\documentclass[twocolumn]{aastex61}

\usepackage{aas_macros}          
\usepackage{nccmath}             
\usepackage{mathtools}           
\usepackage{cases}               
\usepackage{tikz} 
\usepackage{blindtext} 

\newcommand\beq{\begin{equation}}   
\newcommand\eeq{\end{equation}}  
\def\bal#1\eal{\begin{align}#1\end{align}}  
\newcommand\bse{\begin{subequations}}    
\newcommand\ese{\end{subequations}}       

\def\bml#1\eml{\begin{multline}#1\end{multline}}  

\newcommand\bga{\begin{gather}}    
\newcommand\ega{\end{gather}}      

\usepackage{color}    
\newcommand\red[1]{{\color{red}#1}}     
  
\newcommand\rdagger[1]{\red{^{\dagger}}}

\newcommand\kelvin{\, \mathrm{K}}	
\newcommand\second{\, \mathrm{s}}	
\newcommand\persecond{\, \mathrm{s}^{-1} }	
\newcommand\Eday{\,\, \mathrm{Earth}\, \mathrm{days} }	
\newcommand\metre{\, \mathrm{m}}	
	
\newcommand\permetre{\, \mathrm{m}^{-1} }	

\newcommand\unitbar{\, \mathrm{bar}}	
\newcommand\mbar{\, \mathrm{mbar}}	
\newcommand\gauss{\, \mathrm{G}}	

\usepackage{gensymb}     
\usepackage{amsbsy}        
\newcommand\pa{\partial}         						 				
    		
\newcommand\oo[1]{\frac{1}{#1}}      
\newcommand\vect[1]{\mathbf #1}                     
\newcommand\uvect[1]{\widehat{\mathbf #1}}  		
\newcommand\di[1]{\mathrm{#1}}					 	
\newcommand\dd{\mathrm{d}}						 	
\newcommand\ee{\mathrm{e}}						 	

\newcommand\deriv[2]{\frac{\partial #1}{\partial #2} }		
	
\newcommand\oderiv[2]{\frac{\mathrm{d}#1}{\mathrm{d}#2}}

\newcommand\osecderiv[2]{\frac{\mathrm{d}^2#1}{{\mathrm{d}#2}^2}}
\newcommand\lagderiv[1]{\frac{\mathrm{d}#1}{\mathrm{d}t}}	
\newcommand{\e}{\mathrm{e}} 				
\newcommand\textderiv[2]{  \pa {#1} /  \pa {#2}  }	

\newcommand\textoderiv[2]{\mathrm{d}#1/\mathrm{d}#2}
\newcommand\tdrag{\tau_\di{drag} }	
\newcommand\trad{\tau_\di{rad} }
\newcommand\twave{\tau_\di{wave} }

\newcommand\Alf{Alfv\'en~}	

\usepackage{lipsum}

\usepackage{savesym}
\savesymbol{tablenum}
\usepackage{siunitx}
\restoresymbol{SIX}{tablenum}
\usepackage[noabbrev,capitalise]{cleveref}  

\newcommand\sectref[1]{Section \ref{#1}}     
\newcommand\sectsref[2]{Sections \ref{#1} and \ref{#2}}     
\newcommand\sectrefrange[2]{Sections \ref{#1} to \ref{#2}}     
\newcommand\eqnref[1]{Equation (\ref{#1})}     
\newcommand\appref[1]{Appendix \ref{#1}} 

\received{June 26, 2020}
\revised{April 30, 2021}
\accepted{June 22, 2021}
\submitjournal{ApJ}

\shorttitle{The Magnetic Mechanism for Hotspot Reversals in Hot Jupiter Atmospheres}
\shortauthors{Hindle, Bushby, \& Rogers}

\begin{document}

\title{The Magnetic Mechanism for Hotspot Reversals in Hot Jupiter Atmospheres}


\correspondingauthor{Alex Hindle}
\email{alex.hindle@newcastle.ac.uk}

\author[0000-0001-6972-2093]{A. W. Hindle} 
\affil{School of Mathematics, Statistics and Physics, Newcastle University, Newcastle upon Tyne, NE1 7RU, UK}

\author[0000-0002-4691-6757]{P. J. Bushby}
\affil{School of Mathematics, Statistics and Physics, Newcastle University, Newcastle upon Tyne, NE1 7RU, UK}

\author[0000-0002-2306-1362]{T. M. Rogers}
\affil{School of Mathematics, Statistics and Physics, Newcastle University, Newcastle upon Tyne, NE1 7RU, UK}
\affil{Planetary Science Institute, Tucson, AZ 85721, USA }


\begin{abstract}
Magnetically-driven hotspot variations (which are tied to atmospheric wind variations) in hot Jupiters are studied using non-linear numerical simulations of a shallow-water magnetohydrodynamic (SWMHD) system and a linear analysis of equatorial SWMHD waves. In hydrodynamic models, mid-to-high latitude geostrophic circulations are known to cause a net west-to-east equatorial thermal energy transfer, which drives hotspot offsets eastward. We find that a strong toroidal magnetic field can obstruct these energy transporting circulations. This results in winds aligning with the magnetic field and generates westward Lorentz force accelerations in hotspot regions,  ultimately causing westward hotspot offsets. In the subsequent linear analysis we find that this reversal mechanism has an equatorial wave analogy in terms of the planetary scale equatorial magneto-Rossby waves.  We compare our findings to three-dimensional MHD simulations, both quantitively and qualitatively, identifying the link between the mechanics of magnetically-driven hotspot and wind reversals. We use the developed theory to identify physically-motivated reversal criteria, which can be used to place  constraints on the magnetic fields of ultra-hot Jupiters with observed westward hotspots.
\end{abstract}


\section{Introduction} \label{sec:intro}
In recent years the field of exoplanetary research has greatly developed its understanding of exoplanet characterisation both observationally and theoretically. The field has now reached the point where light curves, infrared photometry, and spectra from spaced-based telescopes can be used to test, inform, and update our understanding of the atmospheric dynamics of these closely-orbiting gas giants.  

Generally, observational measurements of hot Jupiters \citep[e.g.,][]{2006Sci...314..623H,2007MNRAS.379..641C,2007Natur.447..183K,2009ApJ...690..822K,2008ApJ...686.1341C,2009ApJ...690L.114S,2010ApJ...723.1436C,2016ApJ...823..122W}, find that these planets have equatorial temperature maxima (hotspots) located eastward of their substellar points. This is consistent with both
hydrodynamic simulations \citep[e.g.,][]{2002A&A...385..166S,2004JAtS...61.2928S,2005ApJ...629L..45C,2006ApJ...649.1048C,2007ApJ...657L.113L,2008ApJ...673..513D,2009ApJ...700..887M,2010ApJ...714.1334R,2010ApJ...710.1395D,2010ApJ...719.1421P,2011MNRAS.413.2380H,2013ApJ...776..134P} and hydrodynamic theory \citep{2011ApJ...738...71S,2020A&A...633A...2D} of synchronously rotating hot Jupiters, which predict that such hotspots are driven eastward by the interaction between mid-to-high latitude geostrophic circulations and equatorial jets. This fundamental behaviour of the hydrodynamic system can also be described in terms of interactions between the system's dominant equatorial waves and its mean equatorial flows \citep{2011ApJ...738...71S}.

However, recent observations suggest eastward hotspots may not be found ubiquitously, particularly on the hottest hot Jupiters (ultra-hot Jupiters). Continuous optical {\em Kepler} measurements find east-west brightspot oscillations on the ultra-hot Jupiters HAT-P-7b \citep{2016NatAs...1E...4A} and Kepler-76b \citep{Jackson_2019}; optical phase curve measurements from {\em TESS} find westward brightspot offsets on the ultra-hot Jupiter WASP-33b \citep{2020arXiv200410767V}\footnote{Although  \cite{2020arXiv200410767V} acknowledge that systematic effects in the data, due to host star variability, cannot be ruled out as a potential cause of their westward brightspot measurements.}; while thermal phase curve measurements from {\em Spitzer} find westward hotspots on the ultra-hot Jupiter WASP-12b \citep{2019MNRAS.489.1995B} and the cooler hot Jupiter CoRoT-2b \citep{2018NatAs...2..220D}.  There are three main explanations for these observations: reflections from cloud asymmetries confounding optical measurements \citep[]{2013ApJ...776L..25D,2016A&A...594A..48L,2016ApJ...828...22P,2017ApJ...850...17R},  asynchronous rotation \citep[]{2014ApJ...790...79R}, and  {magnetism} \citep{2014ApJ...794..132R,2017NatAs...1E.131R,2019ApJ...872L..27H}. Ultra-hot Jupiters generally have near-zero eccentricities and are thought to be tidally-locked, so are expected to be synchronously rotating. They are also expected to have cloud-free daysides, where their atmospheres are too hot for condensates to form. \cite{2019A&A...631A..79H} recently ruled out cloud asymmetries as the explanation for westward brightspots on HAT-P-7b.

Using three-dimensional (3D) magnetohydrodynamic (MHD) studies, \cite{2014ApJ...794..132R} predicted that magnetic fields could cause wind variations that drive east-west hotspot oscillations. \cite{2017NatAs...1E.131R} then showed that the westward venturing hotspot displacements on the ultra-hot Jupiter HAT-P-7b can be well explained by the moderate deep-seated dipolar magnetic field strengths that are expected to be generated in the convective interior of such planets. In \cite{2019ApJ...872L..27H} we used a shallow-water MHD (SWMHD) model to show, firstly, that the magnetically-driven hotspot reversal mechanism is a shallow phenomenon that is driven by the flow's interaction with the planet's atmospheric toroidal magnetic field; and secondly, that the SWMHD model also requires a moderate planetary dipolar magnetic field strength to drive westward hotspot displacements on HAT-P-7b but that an excessively strong deep-seated dipolar magnetic field is required to reverse flows within the cooler (and hence less thermally-ionised) atmosphere of CoRoT-2b. The westward hotspot offsets on CoRoT-2b are therefore more plausibly explained by non-magnetic phenomena. Interestingly, the hot Jupiters Kepler-76b, WASP-12b, and WASP-33b are of the ultra-hot type so are more akin to HAT-P-7b than CoRoT-2b, making magnetically-driven reversals plausible for these observations. 

{While 3D MHD simulations have proved crucial for identifying that magnetism can drive hotspot reversals in ultra-hot Jupiters, their dynamics is often too subtile and complex to glean physical understanding from. The aim of this study is to use a reduced physics model, alongside known features of 3D MHD simulations, to identify the mechanism by which magnetism can reverse hotspots in ultra-hot Jupiters. }

In \sectsref{sect:numeric:model}{sect:numerical:results}, we use the numerical two-layer Cartesian SWMHD model of \cite{2019ApJ...872L..27H}, with an equatorial beta-plane treatment of the Coriolis effect and different purely azimuthal equatorially-antisymmetric initial magnetic field treatments, to study the westward transition of hotspots. 
In \sectref{sect:linear:model}, we examine the link between magnetically-driven wind reversals and equatorial wave dynamics. Finally, in  \sectref{sect:hotspot:reversal:criterion} we collate our findings, compare them to results of 3D MHD simulations, and present a physically-motivated hotspot reversal criterion for ultra-hot Jupiters.

\section{Non-linear shallow-water model} 
\label{sect:numeric:model} 

Three-dimensional models are fundamental to understanding the general features and flow behaviours of planetary atmospheres. However, with so many physical processes in play, it can be difficult to isolate the mechanisms responsible for driving a given flow pattern. In such instances, simplified models can be used to reduce the number of physical processes involved, exposing the underlying physics responsible for specific dynamical features. In this section, we present a detailed description of the reduced-gravity SWMHD model briefly described in \cite{2019ApJ...872L..27H}, which we will use to explore the physics of wind reversals.

\subsection{Governing equations}  \label{subsect:numeric:model}
The reduced-gravity SWMHD model, is an adaptation of the SWMHD model of \cite{2000ApJ...544L..79G} and is the MHD analogue of its hydrodynamic namesake \citep[e.g.,][]{2006aofd.book.....V}, which has been used extensively in hydrodynamic studies of hot Jupiters \citep{2007ApJ...657L.113L,2010GeoRL..3718811S,2013ApJ...762...24S,2013ApJ...776..134P}. It is made up of two constant density fluid layers: a shallow active upper layer and an infinitely-deep inactive lower layer, which has no pressure gradients, velocity fields or induced magnetic fields in the horizontal direction \citep[see][for a model schematic]{2019ApJ...872L..27H}. Physically, the upper layer represents the meteorologically active upper atmosphere and the lower layer represents the deep atmosphere and deep interior of a hot Jupiter. The interface between the two layers is a material surface over which no magnetic flux is permitted to cross. When the system's length scales approach the shallow-water limit (i.e., if typical active layer horizontal scales, $L$, are much larger than the active layer's thickness, $H$), the vertical momentum equation of the full 3D system approaches magneto-hydrostatic balance. This limiting approximation may be used together with the model's interface constraints to vertically integrate the 3D MHD equations over the vertical coordinate, $z$, to yield a shallow-water model with vertically independent variables \citep[see][for further discussion]{2000ApJ...544L..79G,2019ApJ...872L..27H}. Using Cartesian horizontal spatial coordinates, ($x,y$), the dynamical behaviour of the active layer can be described by the following governing equations:
  \beq
  \begin{split}
    \lagderiv{ \vect{u} } + f (\uvect{z} \times  \vect{u})  & = - g \nabla h + ( \vect{B} \cdot \nabla) \vect{B} 
    \\  & \quad - \frac{\vect{u}}{\tdrag}+ \vect{R} + \vect{D}_\nu, 
    \label{eqn:mom}
 \end{split}
  \eeq
  \begin{align}
   \deriv{h}{t} + \nabla \cdot (h \vect{u} )  &= \frac{ h_\di{eq} - h }{\tau_\di{rad}} \equiv Q,   \label{eqn:cont}
  \\
  \lagderiv{ A }  &=  D_\eta, \label{eqn:ind}
  \\
   h \vect{B} &= \nabla \times  A \uvect{z}, \label{eqn:magfluxfunc} 
  \end{align}
where $\vect{u}(x,y,t)\equiv (u,v)$, is the horizontal active layer fluid velocity, $h(x,y,t)$ is the active layer thickness which is used as the model's temperature proxy (see below), $\vect{B}(x,y,t)\equiv (B_x,B_y)$ is the horizontal active layer magnetic field (in velocity units), and $A(x,y,t)$ is the magnetic flux function of the active layer. We comment that the magnetic flux function definition differs from its two-dimensional definition through the inclusion of $h$ in \cref{eqn:magfluxfunc}. This arises as the magnetic flux function describes the vertically-integrated horizontal magnetic field over the whole fluid column, rather than simply the horizontal magnetic field at a specific vertical level. We use $\nabla \equiv (\pa_x, \pa_y)$ to define the horizontal gradient operator, $\di{d}/ \di{dt} \equiv \pa/ \pa t + \vect{u} \cdot \nabla$ to define the Lagrangian time derivative operator, and $\nabla \times  A \uvect{z} \equiv (\pa_y A  , -\pa_x A)$ is the horizontal curl of the scalar field $A$ about the vertical coordinate. 

Defining the system in terms of the magnetic flux function guarantees that the SWMHD divergence-free condition, $\nabla \cdot (h \vect{B}) =0$, remains satisfied throughout the domain at all times. This is the shallow-water analogue of Gauss' law of magnetism, which excludes magnetic monopoles. This shallow-water divergence-free condition is obtained by integrating the full 3D form of Gauss' law over the vertical coordinate, while imposing zero magnetic flux constraints across our model's layer interfaces. Using this formulation also highlights that in the absence of magnetic diffusion ($D_\eta=0$), $A$ is a materially conserved quantity (see \cref{eqn:ind} with $D_\eta=0$). 

For numerical stability, we apply the following explicit diffusion prescriptions \citep[][A.~D.~Gilbert et al.~2021, in preparation]{doi:10.1093/qjmam/hbu004}:
\begin{align}
&\vect{D}_\nu = h^{-1} \nabla \cdot \left[ \nu h \left(\nabla \vect{u} + (\nabla \vect{u})^T \right) \right],
\\
&D_\eta =  \eta (\nabla^2 A-  h^{-1} \nabla h \cdot \nabla A),
\end{align}
where $\nu$ is the kinematic viscosity and $\eta$ is the magnetic diffusivity. 

Geometrically, we fix a local Cartesian coordinate system about the equator, with $-R \pi  \leq x < R \pi$ and $-R\pi /2 < y <R \pi/ 2$.  We centre the system about the planet's substellar point, so $x/R$ approximately corresponds to the azimuthal coordinate and $y/R$ approximately corresponds to the latitudinal coordinate. Rotational effects are included via the so-called {\em equatorial beta-plane} approximation of \cite{Rossby_1939_JMR}. Specifically, the only effects of sphericity the equatorial beta-plane approximation captures are the dynamical effects caused by latitudinal variations in the planetary rotation vector's vertical component. The approximation also uses the fact that in equatorial regions the Coriolis parameter, $f$, is approximately linear to set $f = \beta y$,  where the constant $\beta = 2\Omega / R$ is the local latitudinal variation of the Coriolis parameter at the equator,  $\Omega$ is the planetary rotation rate and $R$ is the planetary radius. 

The system is driven by a Newtonian cooling treatment, $Q$, in the continuity equation (\cref{eqn:cont}), which relaxes the system towards the prescribed radiative equilibrium thickness profile, $h_\di{eq}$, over a radiative timescale, $\trad$. The Newtonian cooling is implemented with 
\beq
    h_\di{eq} =  H + \Delta h_\di{eq} \cos \left( \frac{x}{R} \right) \cos \left( \frac{y}{R} \right), \label{eqn:heqprof} 
\eeq
where $H$ is the system's reference active layer thickness at radiative equilibrium and $\Delta h_\di{eq}$ is the difference in $h_\di{eq}$ between this reference thickness and the radiative equilibrium layer thickness at the substellar point. This profile is similar to the spherical forcing prescriptions used in comparable hydrodynamic models \citep[e.g.,][]{2004JAtS...61.2928S,2007ApJ...657L.113L,2010GeoRL..3718811S,2011ApJ...738...71S,Showman_2012,2013ApJ...776..134P}. The transfer of mass caused by $Q$ generates horizontal pressure gradients, which drive recirculation via the generation of planetary scale shallow-water waves. Similarly, in three dimensional models, pressure gradients caused by heating drive recirculation via internal gravity waves. Using this analogy, mass sources and sinks represent heating and cooling respectively. This connection has been used extensively in hydrodynamic models of hot Jupiters, with active layer geopotential, $gh$, used as a proxy for specific thermal energy \citep{2007ApJ...657L.113L,2010GeoRL..3718811S,2013ApJ...762...24S,2013ApJ...776..134P}. Using this physical link, we equate the model's active layer reference geopotential, $gH$, to the reference thermal energy, $\mathcal{R} T_\di{eq}$, of the modelled planet's atmosphere, where $\mathcal{R}$ and $T_\di{eq}$ respectively denote the specific gas constant and the equilibrium reference temperature.

With the addition of $Q$, a vertical mass transport term, $\vect{R}$, needs to be introduced to enforce specific momentum conservation. In ``cooling'' regions ($Q<0$) mass sinks from the active layer to the quiescent layer and  causes no active layer accelerations\footnote{The momentum that is removed from the active layer is transferred to the quiescent layer. However, since the quiescent layer is infinitely-deep, the momentum of the transferred mass plus the quiescent layer is conserved with no change to the quiescent layer's velocity.}. However, in ``heating'' regions ($Q>0$) mass transport causes deceleration of the active layer as motionless fluid is transferred upwards. This deceleration due to heating is calculated by requiring specific momentum conservation in the active layer, yielding
\begin{equation}
    \vect{R} =
  \begin{cases}
        0  &  \di{for}\,\, Q < 0 \\
        -  \frac{ \vect{u} Q}{h} & \di{for}\,\, Q \geq 0 ,
  \end{cases}
\end{equation}
which has also been used in the hydrodynamic version of this model \citep[e.g.,][]{2004JAtS...61.2928S,2010GeoRL..3718811S,2011ApJ...738...71S,Showman_2012,2013ApJ...776..134P}. 

We parameterise atmospheric drag with a linear Rayleigh drag treatment, $-\vect{u}/\tdrag$, 
where $\tdrag$ is the timescale of the dominant horizontal drag process in the thin active layer. Previous hydrodynamic studies use this Rayleigh drag to parameterise Lorentz forces \citep[e.g.,][]{2010ApJ...719.1421P,2013ApJ...764..103R} or basal drag at the bottom of the radiative zone \citep[e.g.,][]{1994BAMS...75.1825H,2013ApJ...770...42L,2016ApJ...821...16K}. In our study, we include Lorentz forces explicitly. {However, due to the geometry of the SWMHD model, we only explicitly include the Lorentz forces caused by the atmospheric toroidal magnetic field (see \sectref{subsect:initial:Bfield} for a discussion of the magnetic field geometry in the atmosphere)}. We hence use the Rayleigh drag treatment to parameterise the Lorentz forces caused by planet's deep-seated poloidal magnetic field, which are not included explicitly. This is consistent with the treatment proposed by \cite{2010ApJ...719.1421P}, whose $\tdrag$ parameterisation was based on estimating the direct influence that the planet's deep-seated poloidal magnetic field has on zonal flows. Though one could argue that in this setting the Rayleigh drag should have no meridional component, for comparison with past hydrodynamic results, we follow the commonly applied treatment of using Rayleigh drag in both horizontal directions \citep[e.g.,][]{2010ApJ...719.1421P,2011ApJ...738...71S,2013ApJ...764..103R,2013ApJ...776..134P}.\footnote{We find that the meridional component of the Rayleigh drag never has a leading order influence, being 1-2 orders of magnitude smaller than the system's dominant meridional accelerations, so does not qualitatively influence any of our results. An example of this can be seen in \cref{fig:dvdt}. }  We also comment that \cite{2014ApJ...794..132R} found that magnetically driven wind variations emerge in the upper radiative atmosphere (where basal drags are negligible), so we do not consider basal drag in this work.

\subsection{Magnetic field profile} \label{subsect:initial:Bfield}
The extension of planetary dynamo theory into the hot Jupiter regime is not well understood. That said, from current dynamo theory one would expect hot Jupiters to have planetary dynamos that are sustained within the convective deep interior, generating deep-seated poloidal magnetic fields. The hottest hot Jupiters also have weakly-ionised atmospheres. If the atmospheres are sufficiently ionised, the  zonally-dominated atmospheric flows become sufficiently connected to the planet's deep-seated poloidal magnetic field to induce a strong toroidal field that dominates the atmospheric magnetic field geometry  \citep{2012ApJ...745..138M}. Assuming this picture, and the planet's deep-seated magnetic field's geometry is dominated by an axial dipole, the induction of the toroidal component of the magnetic field can be approximated by
\beq
\begin{aligned}
\deriv{\vect{B}_\phi}{t}  & \approx (\vect{B}_\di{dip} \cdot  \nabla_{(3)})  \vect{V}_\phi  -  \nabla_{(3)}  \times (\eta  \nabla_{(3)}  \times \vect{B}_\phi), 
\end{aligned} \label{eqn:3D:induct}
\eeq
where $\vect{B}_\phi \equiv B_\phi \uvect{\phi}$ is the toroidal component of the magnetic field, $\vect{B}_\di{dip}$ is the planetary dipolar field, $\vect{V}_\phi \equiv V_\phi \uvect{\phi}$ is the zonal component of the atmospheric flow, $\nabla_{(3)} $ is the 3D gradient operator, and the electric currents generating the dipolar planetary field are implicitly assumed to be located far below the atmospheric region of interest \citep{2010ApJ...719.1421P,Perna_2010,2011ApJ...738....1B,2012ApJ...745..138M}. Therefore, if toroidal field induction dominates toroidal field diffusion, the atmospheric toroidal field profile is expected to be equatorially-antisymmetric, as found in the simulations of \cite{2014ApJ...794..132R}. 

There are not enough degrees of freedom in the SWMHD induction equation to simultaneously model the planetary dipolar field and the atmospheric toroidal field, so we only model the dominant atmospheric toroidal field self-consistently.  We choose to enforce the simple equatorially-antisymmetric, purely azimuthal, initial magnetic field:
\beq
\vect{B}_0 = B_0 \uvect{x} = V_\di{A} \ee^{1/2} \tanh(y/L_\di{eq}) \uvect{x},  \label{eqn:B0:num}
\eeq
where $V_\di{A}$ is the constant parameter that sets the magnitude of the azimuthal magnetic field. This profile may appear an unintuitive choice at first, but \cite{2017GApFD.111..115L} noted that it has the useful properties for wave dynamics, which we shall exploit in \sectref{sect:linear:model}. It is monotonic, behaves linearly in the equatorial region, and is bounded as $y/L_\di{eq} \rightarrow \infty$. The approximately linear latitudinal dependence of $B_0$ in  the equatorial region means one can choose $V_\di{A}$ in accordance with the first order Taylor expansion of non-monotonic equatorially-antisymmetric profiles. Upon comparing to other field profiles, we generally find that doing so reproduces similar equatorial dynamics. To illustrate this, in \sectref{sect:numerical:results} we compare some basic results to the profile $B_0 = V_\di{A}  (y/L_\di{eq})\exp(1/2-y^2/2 L_\di{eq}^2)$, which is the equatorially-antisymmetric profile used in \cite{2019ApJ...872L..27H}. This has the same first order Taylor expansion as \cref{eqn:B0:num}, has the maximum $B_0=V_\di{A}$ at $y=L_\di{eq}$ (i.e., $V_\di{A}$ is the maximal initial Alfv\'en speed), and can be motivated from both \cref{eqn:3D:induct} and the simulations of \cite{2014ApJ...794..132R}. We implement the initial magnetic field profile of \cref{eqn:B0:num} across an initially flat layer ($h(x,y,0)=H$, everywhere), using the initial magnetic flux function, {$A_0(y) =  H V_\di{A} L_\di{eq} \ee^{1/2} \ln(\cosh(y/ L_\di{eq})))$}.

\subsection{Numerical method and parameter choices} \label{subsect:numeric:method}
{Numerical solutions are obtained by evolving  \cref{eqn:mom,eqn:cont,eqn:ind,eqn:magfluxfunc} from an initial uniformly-flat rest state (i.e., $h(\vect{x},0)=H$, $\vect{u}(\vect{x},0)=\vect{0}$), in the presence of a purely azimuthal magnetic field ($A(\vect{x},0)=A_0(y)$). For hydrodynamic solutions we evolve until steady-state is achieved and for MHD solutions we run for a magnetic diffusion timescale.} The system is solved on a $256 \times 511$ $x$-$y$ grid, using an adaptive third-order Adam-Bashforth time-stepping scheme \citep{2003ApJ...588.1183C}, with spatial derivatives taken pseudo-spectrally in $x$ and using a fourth-order finite difference scheme in $y$. We use periodic boundary conditions on $\vect{u}$, $h$, and $A$ in the $x$ direction.  On the $y$ boundaries we impose $v=0$ (impermeability), $\pa u / \pa y=0$ (stress-free), $\pa A /\pa x = 0$ (no normal magnetic flux), and maintain the total columnar horizontal magnetic flux of the system. These conditions do not fix values of $h$ on the $y$ boundaries, which are updated to satisfy a consistency condition that results from mass conservation and our other boundary conditions.\footnote{The results we present in \sectref{sect:numerical:results} are also robust to other modelling setups, including initialising $h(\vect{x},0)$ and $\vect{u}(\vect{x},0)$ from hydrodynamic steady state profiles and applying different boundary treatments \cite[e.g., extended domains and absorbing boundaries like those discussed in][]{2013imcp.book.....G}. Regardless of these modelling variations, solutions exhibited similar fundamental behaviours (and reversal thresholds).}

 \begin{deluxetable}{ccccc}
\tablecaption{Planetary parameters for HAT-P-7b, where $T_\di{eq}$, $t_\di{orbit}$, $M$, and $R$ respectively denote the equilibrium reference temperature, the orbital period, the planet's mass and the planetary radius. $M_J$ and $R_J$ respectively denote Jupiter's mass and the nominal Jupiter equatorial radius. \label{tab:planet:params}}
\tablecolumns{5}
\tablehead{
\colhead{\hspace{1em} $T_\di{eq} \,(\di{K})$\tablenotemark{$a$}} & 
\colhead{\hspace{1em}$t_\di{orbit}\,(\di{days})$\tablenotemark{$a$}} & 
\colhead{\hspace{1em}$M \,(M_J)$\tablenotemark{$a$}} & 
\colhead{\hspace{1em}$R \,(R_J)$\tablenotemark{$a$} \hspace{1em}}
}
\startdata
2200  & 2.20 & 1.74 & 1.43 \\
\enddata
\tablenotetext{a}{Data taken from \href{www.exoplanet.eu}{www.exoplanet.eu}, accessed June 14, 2020. $T_\di{eq}$ is set to the planet's orbit-averaged effective temperature, as calculated in \cite{Laughlin_2011}, and is given to $2$ significant figures.}
\end{deluxetable}
 
{We choose simulation parameters based on the planetary parameters of HAT-P-7b, an ultra-hot Jupiter with observed east-west brightspot variations \citep{2016NatAs...1E...4A} that can be well explained by 3D MHD simulations \citep{2017NatAs...1E.131R}. Relevant planetary parameters are presented in \cref{tab:planet:params}. As discussed above, we equate the active layer's reference geopotential with a radiative equilibrium thermal energy reference level. Therefore the gravity wave speed is set using $c_g \equiv \sqrt{gH} = \sqrt{\mathcal{R} T_\di{eq}} = \num{3.0e3} \metre \persecond$, where we use the planet's orbit-averaged effective temperature for the equilibrium reference temperature and the specific gas constant is calculated using the solar system abundances in \cite{10.1007/978-3-642-10352-0_8}. We assume synchronous orbits, so $\Omega = 2 \pi / t_\di{orbit}$, where $t_\di{orbit}$ is the orbital period. We calculate $\beta \equiv 2\Omega/R = \num{6.6e-13} \permetre \persecond$, so the equatorial Rossby deformation radius is
\beq
L_\di{eq} \equiv \left( \frac{c_g}{\beta} \right)^{1/2} \approx \num{6.7e+07} \metre. \label{eqn:Leq}
\eeq
This is a fundamental length scale over which gravitational and rotational effects balance, and is the interaction length scale of planetary scale flows that corresponds to their latitudinal widths. 

The characteristic wave travel timescale, $\tau_\di{wave}$, is defined by the time a shallow-water gravity wave takes to travel over the distance $L_\di{eq}$, and is
\beq 
\begin{aligned}
\tau_\di{wave} \equiv \frac{L_\di{eq}}{c_g} & \approx \num{2.2e+04} \second  \approx 0.26 \Eday,
\end{aligned}
\eeq
We set the reference thickness of the model's active layer to the atmospheric pressure scale height, that is $H \equiv {\mathcal{R} T_\mathrm{eq}R^2}/{G M} = \num{4.3e5} \metre$, where $M$ is the planetary mass and $G$ is Newton's gravitational constant. }

In hydrodynamic shallow-water models \citep[e.g.,][]{2004JAtS...61.2928S,2007ApJ...657L.113L,2010GeoRL..3718811S,2011ApJ...738...71S,Showman_2012,2013ApJ...776..134P}, the forcing profile is usually set so that $\Delta h_\di{eq}/H \sim (T_\di{day}-T_\di{eq})/T_\di{eq}$, where $T_\di{eq}$ is the average reference temperature (for a given atmospheric depth) and $T_\di{day}$ is the maximal dayside reference temperature (at that atmospheric depth). For comparison, applying the reference temperatures used for HAT-P-7b in \cite{2017NatAs...1E.131R}, this equates to $\Delta h_\di{eq}/H \sim 0.22$, $0.19$, and  $0.14$ at $P = 10^{-3}\unitbar$, $10^{-2}\unitbar$, and $10^{-1} \unitbar$ respectively.  {We consider models with {$\Delta h_\di{eq}/H = \{0.01, 0.05, 0.1, 0.2, 0.3, 0.4, 0.5, 0.6 \}$} to cover forcing parameter regimes within and either side of the expected range. }

The simulations presented in this paper have a viscous diffusion of {$\nu = \num{4e8} \metre^2 \persecond$}. In terms of ``true'' physical values, this diffusion coefficient is comparatively large; yet, upon checking, we find that viscous components of \cref{eqn:mom} remain negligibly small. This is to be expected as we are predominantly modelling large scale planetary flows, upon which viscous dissipation generally has little direct influence. We set the magnetic diffusivity to {$\eta = \num{4e8} \metre^2 \persecond$, which within the expected $\eta$ range on HAT-P-7b's nightside}. These values of $\eta$ and $\nu$ are both small enough to make the dynamical timescales of our system much smaller than the diffusion timescales. In 3D geometries, longitudinal variations in $\eta$ are likely to play an important role in the evolution of the magnetic field,  but we defer considerations of this more complicated problem to a future paper.

 {The timescales $\trad$ and $\tdrag$ respectively determine the frequency over which Newtonian cooling 
 and magnetic drag from the deep-seated (but not atmospheric) magnetic field are allowed to occur. Studies of hydrodynamic shallow-water analytics \citep{2011ApJ...738...71S} and simulations \citep{2013ApJ...776..134P} show that varying $\trad$ controls the efficiency of (geopotential) energy redistribution occurs; whereas varying $\tdrag$ adjusts the distance over which atmospheric re-circulation patterns can flow before becoming significantly damped. Hence, since $\trad$ and $\tdrag$ adjust qualitatively similar (albeit non-identical) fundamental flow features, it can be beneficial to reduce the modelling problem by fixing $\tdrag=\trad$. We do so in three of the examined cases of \sectref{sect:numerical:results}: $(a)$ short $\trad$ and strong drag, $\trad=\tdrag=\twave$; $(b)$ moderate $\trad$ and moderate drag, $\trad=\tdrag=5\twave$; and $(c)$ long $\trad$ radiative and weak drag, $\trad=\tdrag=25\twave$. However, as $\trad$ and $\tdrag$ are not necessarily  equivalent in hot Jupiter atmospheres, we also consider the additional two cases: $(d)$ short $\trad$ and weak drag, $\trad=\twave, \tdrag=25\twave$;  and $(e)$ long $\trad$ and strong drag, $\trad=25\twave, \tdrag=\twave$. \cite{2014ApJ...794..132R} found magnetically-driven reversals to occur in the upper atmospheres of ultra-hot Jupiters, where $\trad\sim \twave$ and $\tdrag \sim \twave$, the conditions are most akin to case $(a)$, though $\trad$ and $\tdrag$ are not generally exactly equal.  }

The remaining free parameter in our system is {$V_\di{A}$}, which determines the magnitude of the system's magnetic field. Our general approach is to increase {$V_\di{A}$}, from $V_\di{A}=0$, until we find a change in the nature of the SWMHD system (i.e., hotspot reversals). Here we highlight that, for large enough {$V_\di{A}$}, we always find hotspot reversals in the SWMHD model, regardless of our choices of $\Delta h_\di{eq}/H$, $\trad$, and $\tdrag$. In \sectref{sect:numerical:results} we will discuss both hydrodynamic and magnetohydrodynamic solutions over a wide range of parameter choices to illustrate the magnetic mechanism that drives reversals, and its robustness to changes in parameter space.

\subsection{Model validity}
{Here we briefly discuss validity criteria for our model in the context of our parameter choices. First, we comment that $H/L_\di{eq} \simeq \num{6e-3} \ll 1$, so the shallow-water approximation is well-founded and vertical dependences in the atmosphere are not of leading order importance. Secondly, we take $\vect{\Omega}=\Omega \uvect{z}$, which is typically known as the traditional approximation and is formally valid in the limit of strongly stable stratification \citep[$N^2/\Omega^2\gg1$, e.g.,][]{2006aofd.book.....V}. For our parameters, $N^2/\Omega^2 \sim 4\times10^4 \gg1$, so this approximation is also well-founded. Thirdly, our $y$ boundaries are located at $y=\pm R\pi/2 \sim \pm 2.3 L_\di{eq}$, so the impermeable wall at our model's ``poles'' has little physical influence on our solutions and does not interact with the equatorial dynamics we wish to study. Finally, the equatorial beta-plane truncation of the Coriolis parameter is $f = \beta y + O((y/R)^3)$, so we are careful not to draw conclusions about the polar flows (with $y\gtrsim R$), where the Coriolis parameter is overestimated and boundary effects can occur. The initial magnetic field choice of \cref{eqn:B0:num} is based on an analogous Taylor truncation, so places no further constraint on our discussion.}

\begin{figure*}
\centering
\includegraphics[width=0.8\linewidth]{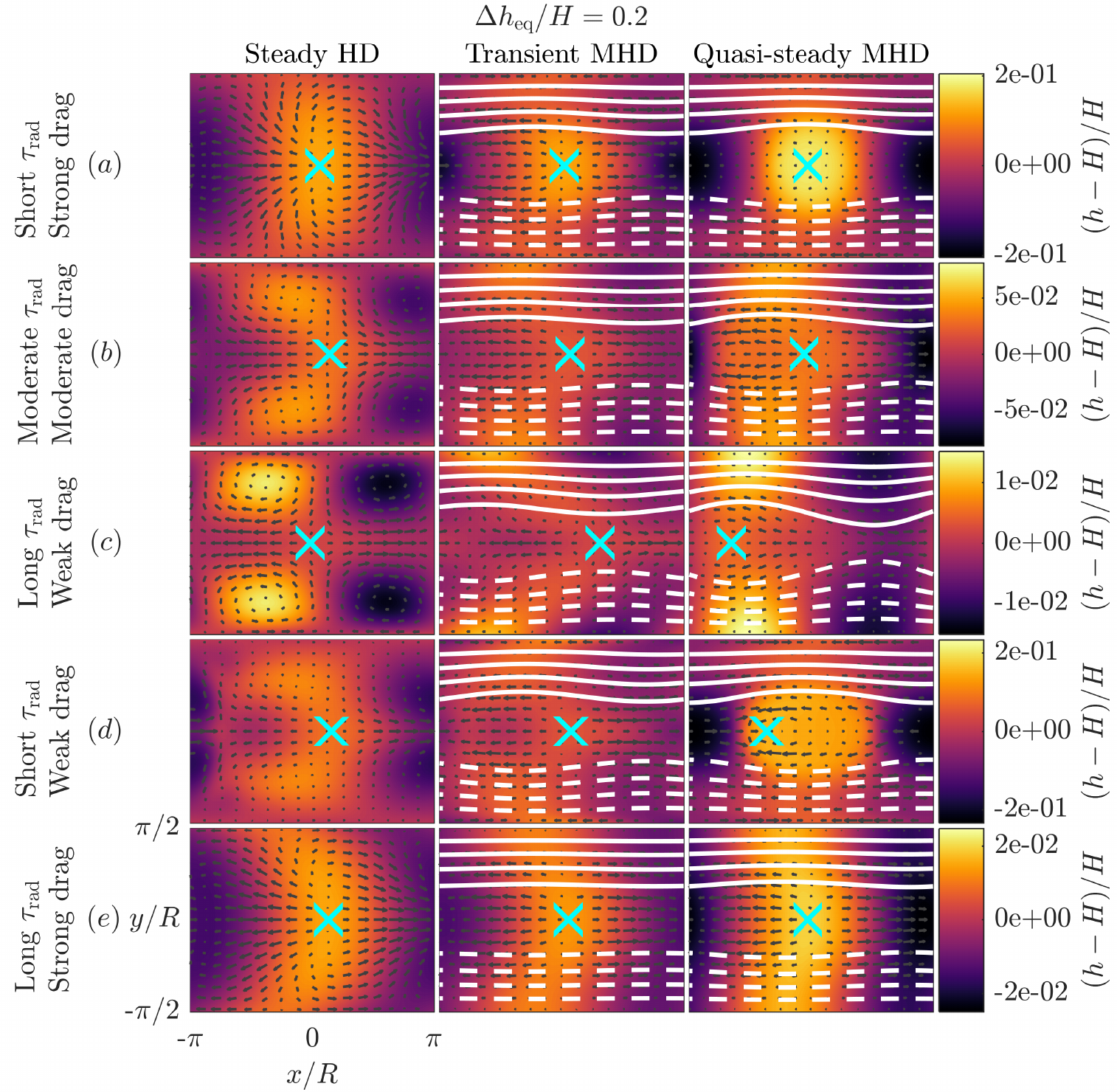} %
\caption{The effect of azimuthal magnetic fields on energy redistribution. Contours of the relative layer thickness deviations (rescaled geopotential energy deviations) are plotted on colour axes that are shared along rows, with (individually-normalised) velocity vectors, hotspots (cyan crosses), and lines of constant $A$ (solid/dashed for $B_x$ positive/negative) over-plotted. In each column, reading from left to right, we present hydrodynamic steady state solutions ($V_\di{A}=0$), supercritical MHD solutions moments before reversal, and supercritical MHD solutions in the reversed quasi-steady phase. {We present solutions in the following parameter regimes: $(a)$ $\trad = \tdrag = \twave$, with $V_\di{A}=0$ or $V_\di{A}=1.6 c_g$ in the top row; $(b)$ $\trad = \tdrag = 5\twave$, with $V_\di{A}=0$  or $V_\di{A}=0.7 c_g$  in the second row;  $(c)$ $\trad = \tdrag = 25 \twave$, with $V_\di{A}=0$  or $V_\di{A}=0.2 c_g$   in the third row;  $(d)$ $\trad = \twave, \tdrag = 25 \twave$, with $V_\di{A}=0$  or $V_\di{A}=1.4 c_g$   in the fourth row;  $(e)$ $\trad = 25 \twave, \tdrag = \twave$, with $V_\di{A}=0$  or $V_\di{A}=0.5 c_g$  in the bottom row.}} \label{fig:Geop}
\end{figure*}

\section{Numerical solutions} \label{sect:numerical:results}
In this section we discuss numerical solutions of the model presented in \sectref{sect:numeric:model}. {First, in \sectsref{subsect:basic:HD}{subsect:basic:MHD} we respectively highlight the basic flow behaviours of  hydrodynamic and magnetohydrodynamic solutions. Then, in \sectref{subsect:Force:balances}, we discuss detailed force balances of these numerical solutions. In \sectrefrange{subsect:basic:HD}{subsect:Force:balances}, we focus on solutions with $\Delta h_\di{eq}/H=0.2$, which lies within the expected forcing range of our fiducial planet HAT-P-7b (see \sectref{sect:numeric:model}). Finally, in \sectsref{subsect:Forcing:dependence}{subsect:B0:comparisons}, we discuss the extension of the developed theory to other forcing magnitudes and toroidal field profiles. }


We visualise the basic form of our numerical solutions by plotting their (non-dimensionalised) geopotential distributions in \cref{fig:Geop}.  As discussed in \sectref{sect:numeric:model}, we use geopotential energy, $gh$, as a shallow-water proxy of thermal energy so the geopotential distributions are analogous to those of temperature perturbations. In the hydrodynamic version of our shallow-water model, solutions are known to converge upon a steady state  \citep[e.g.,][]{2007ApJ...657L.113L,2010GeoRL..3718811S,2013ApJ...762...24S,2013ApJ...776..134P} and we replicate such hydrodynamic steady state solutions in the lefthand column of \cref{fig:Geop} for comparison with our MHD simulations, which we plot in the middle and righthand columns for two difference solution phases (see \sectref{subsect:basic:MHD}). {In each row of \cref{fig:Geop} (from top to bottom) we display the solutions for $(a)$ short $\trad$ and strong drag, $\trad=\tdrag=\twave$; $(b)$ moderate $\trad$ and moderate drag, $\trad=\tdrag=5\twave$; $(c)$ long $\trad$ and weak drag, $\trad=\tdrag=25\twave$;  $(d)$ short $\trad$ and weak drag $\trad=\twave, \tdrag=25\twave$; and $(e)$ long $\trad$ and strong drag $\trad=25\twave, \tdrag=\twave$.} 

\subsection{Basic hydrodynamic solutions} \label{subsect:basic:HD}
Generally, in the hydrodynamic steady state solutions (\cref{fig:Geop}, lefthand column) there are two dominant flow features. Drag-adjusted geostrophic circulations dominate at mid-to-high latitudes; while zonal jets dominate at the equator.  The drag-adjusted geostrophic circulations satisfy a three-way force balance between horizontal pressure gradients, the Coriolis force, and Rayleigh drag (see \sectref{subsect:Force:balances}). In the northern hemisphere, this balance is characterised by flows that circulate clockwise about the geopotential maximum and anticlockwise about the geopotential minimum; while the converse is true in the southern hemisphere. {The dominant acceleration components in the equatorial regions are horizontal pressure gradients, which are largest in the zonal direction; the Rayleigh drag, which is simply a damping force that reduces wind speeds; {and an advection correction, which is of lower order importance if drags are not weak} (again, see \sectref{subsect:Force:balances})}. Hotspots are, by definition, located at the equatorial pressure maxima so the  pressure driven zonally-directed equatorial jets diverge from them. 

{Newtonian cooling drives a solution's geopotential distribution towards the equilibrium geopotential (see that $gh\rightarrow gh_\di{eq}$ as $\trad \rightarrow 0$). Therefore $\trad$ determines two things: how far planetary flows can redistribute geopotential energy before cooling occurs; and the magnitude of pressure gradients in the system, which in-turn determine planetary flows magnitudes (see \cref{fig:Geop}, lefthand column and axis scales). The Rayleigh drag reduces wind speeds everywhere. At equatorial latitudes, a strong Rayleigh drag decreases the distance that the zonal jets can redistribute geopotential energy along the equator, increasing the relative severity of zonal geopotential gradients. At mid-to-high latitudes the Coriolis force becomes significant and solutions satisfy the aforementioned drag-adjusted geostrophic balance. In a ``true'' geostrophic balance, without suppression from drags and forcing, pressure gradients are exactly balanced by the Coriolis force, which acts perpendicularly to the velocity causing flows to rotate (to their right in the northern hemisphere and to their left in the southern hemisphere).  This yields  large-scale mid-to-high latitude vortices that are aligned with isobars, similar to those seen in the short $\trad$, weak drag, hydrodynamic solution (\cref{fig:Geop} $(c)$, lefthand column).  However, the slowing of winds from the Rayleigh drag reduces the magnitude of Coriolis deflection. Therefore in the strong drag limit large-scale vortices cannot fully develop. Similarly, when $\trad$ is short, heating/cooling occurs before large-scale vortices fully develop. Comparing the mid-to-high latitude flows of the hydrodynamic solutions, one finds a transition between the long-$\trad$/weak-drag solutions, with fully-formed geostrophic vortices, to the short-$\trad$/strong-drag solutions, in which the drag-adjusted geostrophic circulations are approximately aligned with the isobars of the equilibrium geopotential (see \cref{fig:Geop}, lefthand column). Aside from an unphysical special case discussed in \cite{2011ApJ...738...71S} and \cite{2013ApJ...776..134P}, for all finite physically-relevant choices of $\trad$ and $\tdrag$, the meridional mass transport into the equator, caused by the drag-adjusted geostrophic circulations, is maximised
east of the substellar point.}

These solutions always exhibit eastward hotspots. This is because the equatorward (rescaled) geopotential energy transport from the mid-to-high latitude circulations, $-\pa(hv)/\pa y$, always has its equatorial maximum located eastward of the substellar point. At the equator, the pressure gradient drives winds that diverge from hotspots, causing equatorial geopotential energy transport away from the hotspot regions (i.e., $-\pa(hu)/\pa x<0$ in hotspot regions). Hence, by \cref{eqn:cont} (geopotential energy conservation), the hotspots locate themselves at the equatorial point of maximal incoming geopotential energy flux, which is located between the equatorial maxima of $-\pa(hv)/\pa y$ and $Q$. The Newtonian cooling ($Q$) attempts to return a solution to its forcing equilibrium (i.e., with its hotspot at the substellar point); whereas, as stated above, the equatorial maximum of $-\pa(hv)/\pa y$ is always eastward. The degree of the hotspot's eastward offset is therefore determined by the location of the equatorial maximum of $-\pa(hv)/\pa y$ and its relative magnitude compared to $Q$. In short, the size of the (eastward) hotspot offset is determined by the efficiency over which the drag-adjusted geostrophic circulations can redistribute thermal\footnote{Recall that the geopotential potential energy is a proxy for thermal energy in this model.} energy from the western equatorial dayside to the eastern equatorial dayside, by circulating it to-and-from the higher latitudes.

\subsection{Basic magnetohydrodynamic solutions} \label{subsect:basic:MHD}

In the weakly-magnetic limit, shallow-water magnetohydrodynamic solutions behave much like their hydrodynamic counterparts (i.e., solutions reach a steady state that is characterised by eastward hotspots, zonal equatorial winds, and drag-adjusted geostrophic circulations at mid-to-high latitudes). However, when the azimuthal magnetic field exceeds a critical magnitude the nature of the solution changes. {\em Supercritical} magnetic solutions have three phases: an {\em initial phase}, in which winds and geopotentials resemble their hydrodynamic counterparts but their circulations induce magnetic field evolution; a {\em  transient phase}, in which mid-to-high latitude winds align with the azimuthal magnetic field and dayside equatorial winds experience a net westward acceleration, driving an east-to-west hotspot transition; and a reversed {\em quasi-steady phase}, in which westward zonally-dominated dayside winds maintain westward hotspots (until, after a comparably long period of time, the magnetic field decays via magnetic diffusion).\footnote{Typically, for these parameters, $\tau_\di{dyn}/\tau_\eta  \sim 0.01$-$0.1$, where $\tau_\di{dyn}$ is the dynamical timescale of the hotspot transition and $\tau_\eta = L_\di{eq}^2/\eta$ is the magnetic diffusion timescale.}  
We present geopotential distributions of supercritical magnetic solutions in the transient and quasi-steady phases in the two righthand columns of \cref{fig:Geop} (middle and right respectively). The supercritical magnetic  solutions are plotted for the same drag choices as the hydrodynamic solutions that they share a row with (see \sectref{subsect:basic:HD}), but now lines of constant $A$, which approximately correspond to field lines of the horizontal magnetic field, are also over-plotted for visualisation of the magnetic field.

After a magnetic solution's initial phase, in which it behaves similarly to its hydrodynamic counterpart, in mid-to-high latitude regions there is a competition between the drag-adjusted geostrophic balance and the magnetic tension (i.e., $\vect{B}\cdot \nabla \vect{B} $, the restorative force that acts to straighten bent horizontal magnetic field lines) that the circulating flows generate. Initially, the magnetic field is purely azimuthal, with only latitudinal gradients in its profile, so magnetic tension is zero everywhere. To understand the magnetic field's evolution we highlight that, as the magnetic diffusion timescale is large in comparison to the dynamical timescales of the system, $A$ is approximately materially conserved.  This means that lines of constant $A$ are advected by the mid-to-high latitude circulations, bending them and causing a growth of magnetic tension. For subcritical magnetic field strengths, a drag-adjusted magneto-geostrophic balance can be supported, with winds and geopotential profiles making small adjustments to balance the magnetic contribution (before magnetic diffusion eventually returns the system to a hydrodynamic steady state). In contrast, for supercritical magnetic field strengths, magnetic tension becomes strong enough to obstruct the drag-adjusted geostrophic circulations and solutions enter into a transient phase, which ultimately results in hotspot reversals. {In \sectref{subsect:Force:balances}, we shall see that the reversal is driven by a westward Lorentz force acceleration in the region surrounding the hotspot, which is itself generated by this obstruction of geostrophic balance. The westward Lorentz force acceleration causes the point of zonal wind divergence on the equator to shift eastwards, so that in hotspot regions geopotential energy flux is westward (i.e., $gh u<0$) rather than zero. This shifts the hotspot westward until the system rebalances into a state with a westward hotspot (again, see \sectref{subsect:Force:balances}).}

{We find that this reversal mechanism (i.e., westward equatorial-dayside Lorentz force accelerations driven by the obstruction of geostrophic balance) always leads to hotspot reversals in the SWMHD model, regardless of our choice of $\Delta h_\di{eq}/H$, $\trad$, and $\tdrag$. However, since these parameters control pressure gradient magnitudes and recirculation efficiency, they determine the critical magnetic field strength sufficient for reversal. We present bounds on the magnetic field strength's critical magnitude, $V_\di{A, crit}$, for various parameter choices in \cref{fig:VAcrit:numeric}. Generally, $\Delta h_\di{eq}/H$ and $\trad$ set the magnitude of a solution's pressure gradients, and therefore the magnitude of the circulations to be overcome, so shorter $\trad$ and larger $\Delta h_\di{eq}/H$ correspond to larger $V_\di{A, crit}$ magnitudes. Initially in long $\tdrag$ solutions the fully formed large scale geostrophic vortices advect the lines of constant $A$ efficiently until they are resisted by magnetic tension; whereas, for short $\tdrag$ solutions, the slowing of winds from drags decreases the distance over which winds initially advect the lines of constant $A$. Therefore
 weak drag solutions generally experience a larger degree of field line bending and hence more magnetic tension (relative to the other accelerations in their solutions for a given $V_\di{A}$) than strong drag solutions. Put simply, strong drag solutions require larger $V_\di{A, crit}$ magnitude to reverse. We quantify dependences of  $V_\di{A, crit}$ on  $\Delta h_\di{eq}/H$, $\trad$, and $\tdrag$ in later discussion.

 In the quasi-steady phase of supercritical SWMHD solutions, the magnitudes of $\trad$ and $\tdrag$ determine the efficiency of the westward energy redistribution. For large $\trad$ and $\tdrag$ timescales,  the (westward) hotspot offsets are large as the equatorial pressure-Lorentz balance is free to redistribute energy towards the point where the zonal winds converge, almost entirely without restriction;  Conversely, for short $\trad$ and $\tdrag$ timescales, this equatorial energy redistribution is less efficient and hotspot offsets are smaller. Comparing between rows in \cref{fig:Geop} (righthand column), suggests $\tdrag$ is the most influential timescale in determining westward hotspot offsets in the SWMHD system.}

\begin{figure*}
\centering
\includegraphics[scale=1]{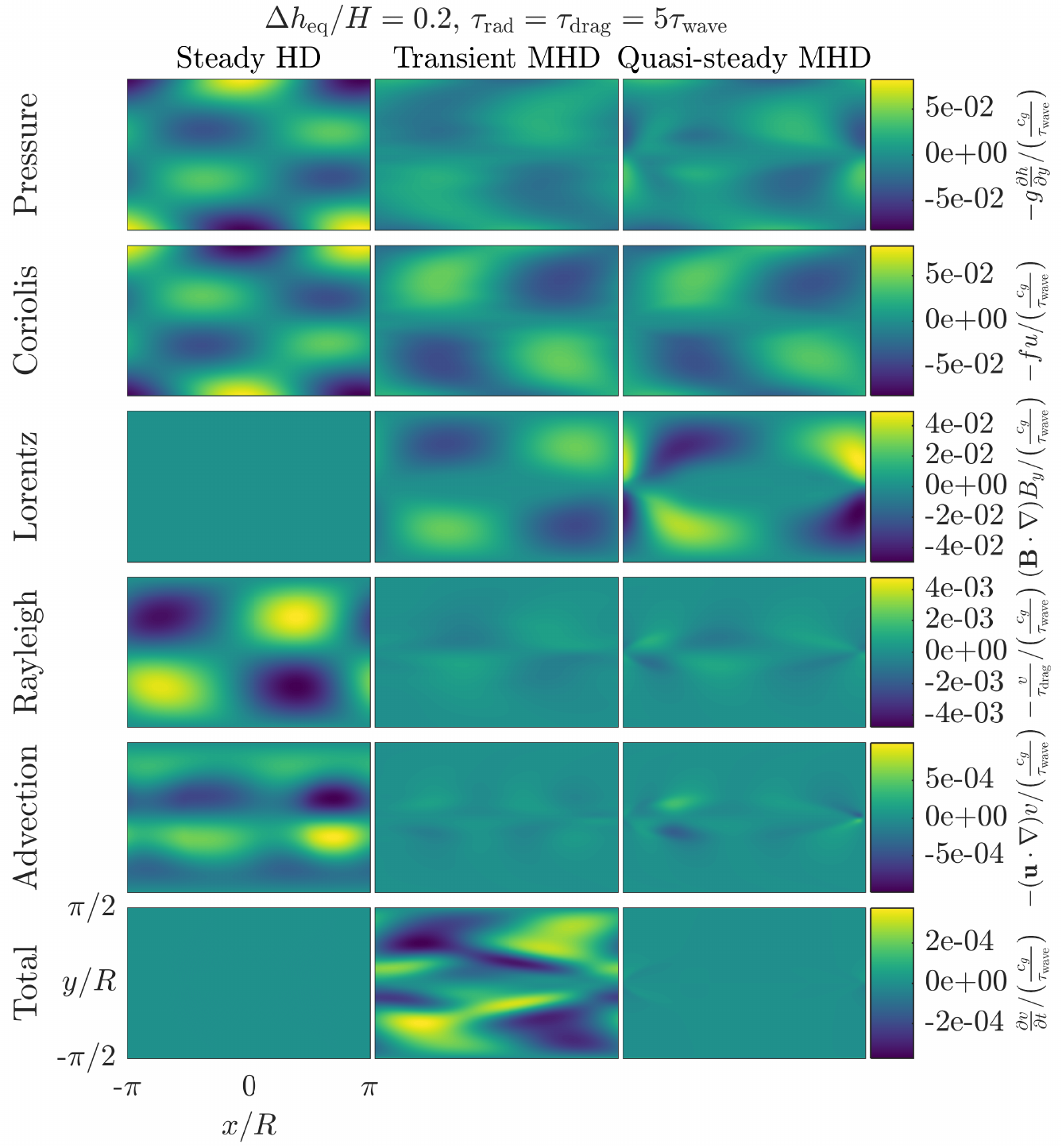} 
\caption{Meridional force balances. In each column, reading from left to right, we plot meridional accelerations corresponding to hydrodynamic steady state solutions, transient phase supercritical MHD solutions, and quasi-steady supercritical MHD solutions. In rows one to four, we respectively plot meridional accelerations due to horizontal pressure gradients, the Coriolis effect, the Lorentz force, and Rayleigh drag; the summed meridional accelerations are plotted in row five. {The solutions are presented for $\Delta h_\di{eq}/H=0.2$, $\trad = \tdrag = 5\twave$, with $V_\di{A}=0$ (HD)  or $V_\di{A}=0.7 c_g$ (MHD)} (i.e., parameter regime $(b)$ in \cref{fig:Geop}).}  \label{fig:dvdt}
\end{figure*}

\subsection{Force balances} \label{subsect:Force:balances}
{In this subsection we compare the force balances of \cref{eqn:mom} for hydrodynamic and supercritical MHD solutions with the parameters of regime $(b)$ in \cref{fig:Geop}  (i.e.,  for $\Delta h_\di{eq}/H=0.2$, $\trad = \tdrag = 5\twave$, with either $V_\di{A} = 0$ or  $V_\di{A} = 0.7c_g$).  We highlight how the presence of a strong equatorially-antisymmetric azimuthal magnetic field modifies the force balances of different planetary regions, and link these modifications to the more general discussions of \sectsref{subsect:basic:HD}{subsect:basic:MHD}.}

\begin{figure*}
\centering
\includegraphics[scale=1]{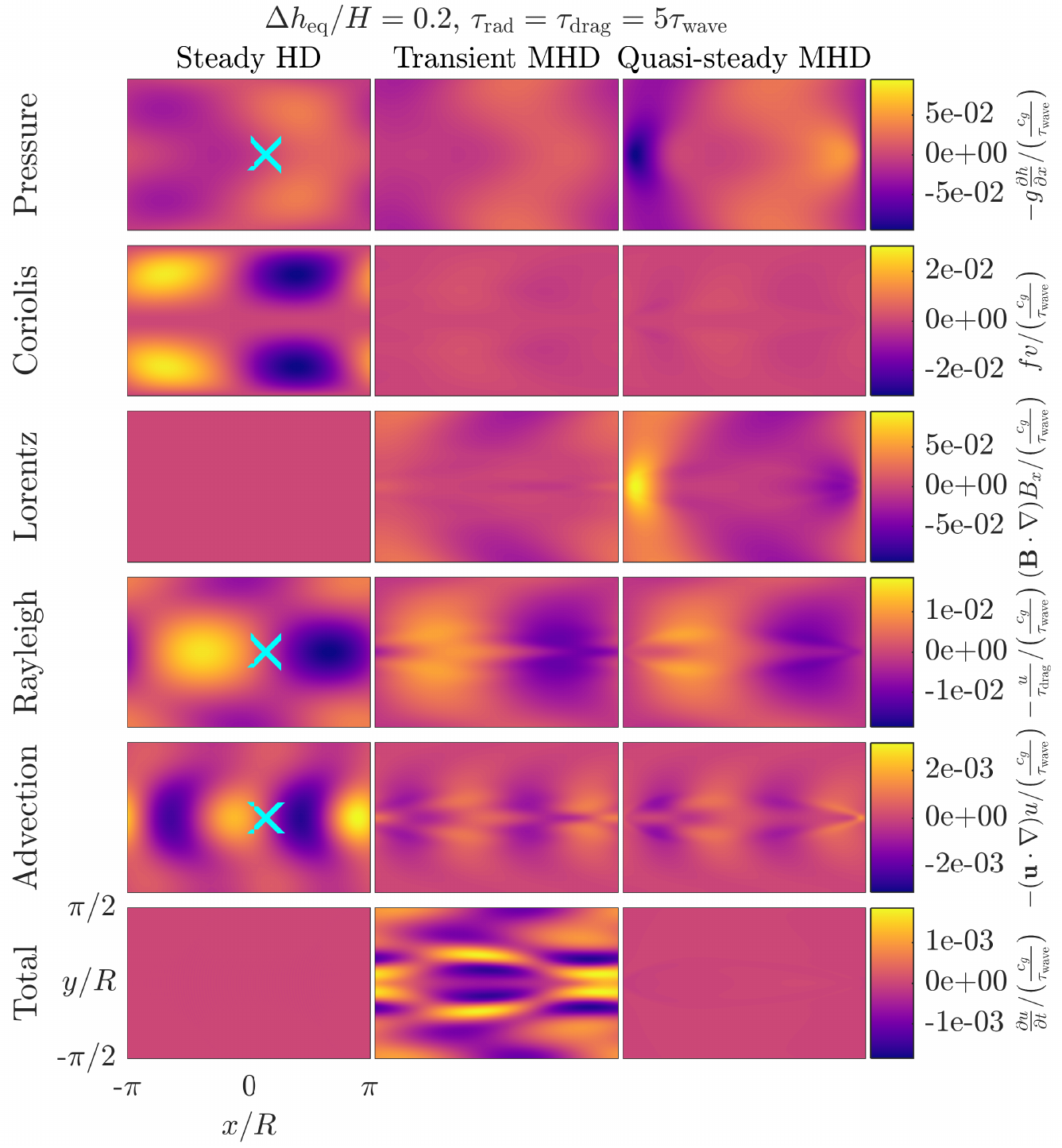} 
\caption{The zonal force balances corresponding to the meridional force balances of \cref{fig:dvdt} (see \cref{fig:dvdt}  caption). As in \cref{fig:dvdt}, we present solutions for the {parameter choices  $\Delta h_\di{eq}/H=0.2$, $\trad = \tdrag = 5\twave$, with $V_\di{A}=0$ (HD)  or $V_\di{A}=0.7 c_g$ (MHD)}  (i.e.,  parameter regime $(b)$ in \cref{fig:Geop}). To aid discussion in the text, hotspot locations have been marked with cyan crosses in hydrodynamic solution panels that correspond to zonal acceleration components with a non-zero equatorial contribution. } 
\label{fig:dudt}
\end{figure*}

In \cref{fig:dvdt,fig:dudt} we respectively plot the dominant meridional and zonal acceleration components of \cref{eqn:mom}, for solutions in regime $(b)$. In the lefthand column of \cref{fig:dvdt,fig:dudt}, we present the acceleration components for the hydrodynamic steady state solution; whereas in the middle and righthand columns of \cref{fig:dvdt,fig:dudt}, we present the acceleration components of the transient and quasi-steady phases of its supercritical MHD counterpart. Along each row of Figures \ref{fig:dvdt} (meridional components) and \ref{fig:dudt} (zonal components), we plot (from top downwards) the acceleration contributions due to horizontal pressure gradients ($-g\nabla h$), the Coriolis effect ($-f\, \uvect{z}\times \vect{u}$), the Lorentz force ($\vect{B}\cdot \nabla \vect{B}$), Rayleigh drag ($-\vect{u}/\tdrag$), and advection ($-\vect{u}\cdot \nabla \vect{u}$). Additionally, in the bottom row of \cref{fig:dvdt} we plot the total meridional acceleration ($\pa v / \pa t$) and, likewise, in the bottom row of \cref{fig:dudt} we plot the total zonal acceleration ($\pa u / \pa t$). For the presented parameter choices  the acceleration contributions due vertical mass transport ($\vect{R}$) and viscous diffusion ($\vect{D}_\nu$) are much weaker so are not included in the plots.

At mid-to-high latitudes, the force balances of hydrodynamic solutions in steady state are well described by the three-way drag-adjusted geostrophic balance discussed in \sectref{subsect:basic:HD}. In particular, \cref{fig:dvdt,fig:dudt} (lefthand column) highlight this for regime $(b)$, showing that in both horizontal directions the mid-to-high latitude accelerations due to horizontal pressure gradients and the Coriolis force almost exactly cancel, albeit with small Rayleigh drag adjustment and a {yet smaller advection contribution}. The meridional components of these accelerations remain balanced in equatorial regions, with all of them vanishing at the equator. However, in the zonal direction, the Coriolis force vanishes in equatorial regions but zonally-directed pressure gradients do not, so zonal pressure gradients are balanced by the Rayleigh drag, {with an advection adjustment}. Since hotspots in hydrodynamic solutions are always located where zonal equatorial jets diverge, these three acceleration components are equally zero at hotspots (see cyan markers in \cref{fig:dudt}).  As discussed in \sectref{subsect:basic:HD}, hotspots are driven eastward by the net west-to-east equatorial energy transfer that results from the mid-to-high latitude drag-adjusted geostrophic circulations.

As discussed in \sectref{subsect:basic:MHD}, magnetic tension ($\vect{B}\cdot \nabla \vect{B}$) is initially zero everywhere so MHD solutions initially resemble their hydrodynamic counterparts. {However, lines of constant $A$ (which closely follow magnetic field lines) are advected by the mid-to-high latitude circulations that are archetypal of hydrodynamic solutions.} This causes them to bend equatorward between the western and eastern dayside (where the initial circulations are poleward and equatorward respectively; see \cref{fig:Geop}, row $(b)$, middle column). Consequently, a restorative Lorentz force that resists meridional winds is produced (see \cref{fig:dvdt}, third row, middle column).  For subcritical MHD solutions (not plotted) this Lorentz force resists but does not fully obstruct the mid-to-high latitude circulations, which adjust into a (drag-adjusted) magneto-geostrophic balance. However, in supercritical MHD solutions, the Lorentz force resists meridional winds strongly enough to zonally-align the mid-to-high latitude winds. Hence, supercritical MHD solutions enter into the transient phase discussed in \sectref{subsect:basic:MHD}.

\begin{figure*}
\centering
\includegraphics[scale=1]{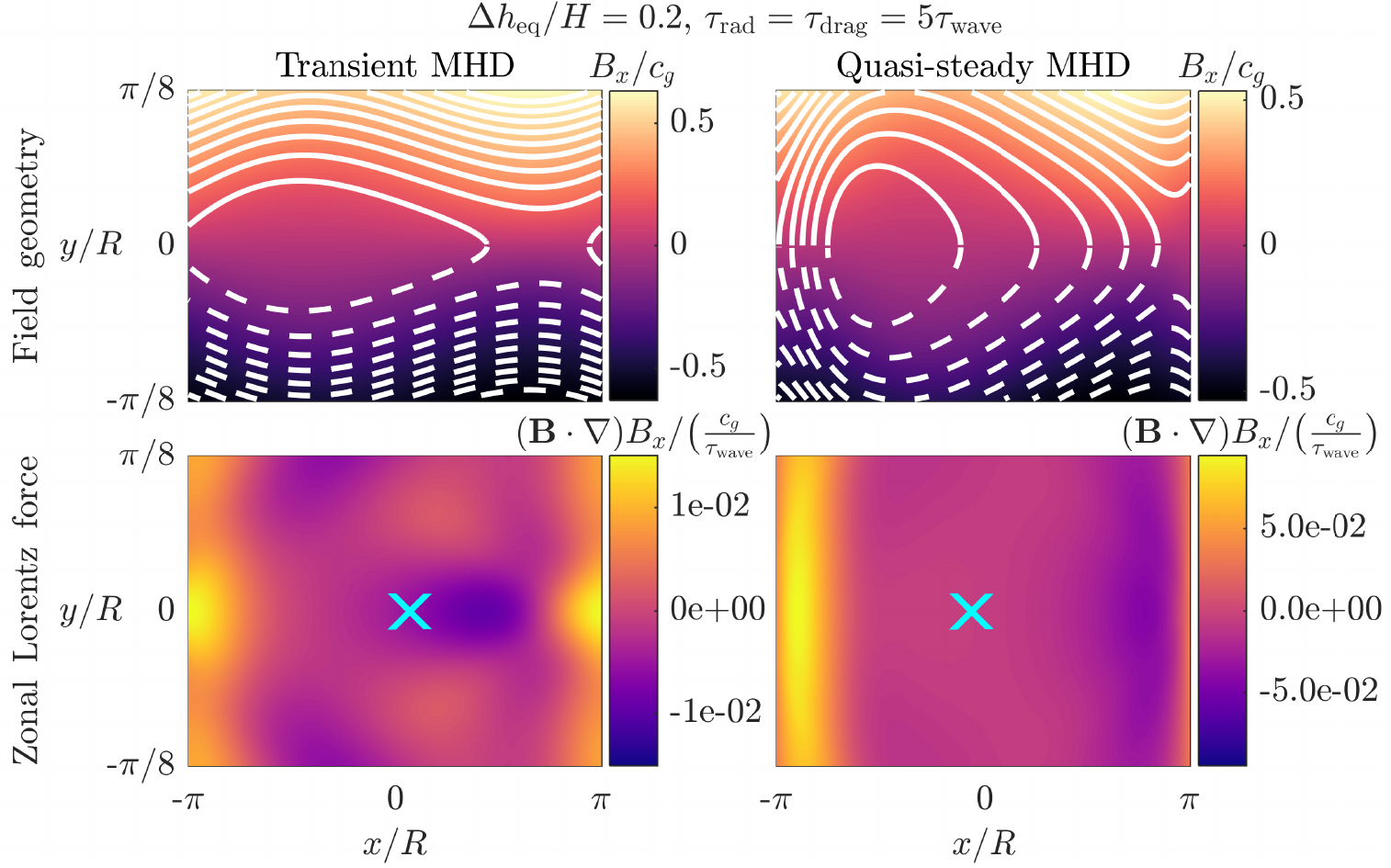} 
\caption{The Lorentz force drives westward accelerations in hotspot (cyan crosses) regions. The azimuthal component of the magnetic field is plotted in the top row, with contours of constant $A$ overlaid (white solid/dashed contours for $B_x$ positive/negative). The corresponding zonal Lorentz force component is plotted in the bottom row. As in the two righthand columns of \cref{fig:dvdt,fig:dudt}, we present the transient (lefthand column) and quasi-steady (righthand column) phases of the supercritical MHD solution with  {$\Delta h_\di{eq}/H=0.2$, $\trad = \tdrag = 5\twave$, and $V_\di{A}=0.7 c_g$} (i.e.,  parameter regime $(b)$ in \cref{fig:Geop}), though we restrict this plot to the equatorial region, $-\pi/8<y/R<\pi/8$. The zonal Lorentz force acceleration is the directional derivative of $B_x$ along horizontal magnetic field lines (approximately lines of constant $A$, see \sectref{subsect:numeric:method}).} 
\label{fig:BxEq}
\end{figure*}

When the magnetic field geometry is azimuthally dominated, understanding Lorentz force accelerations is less intuitive in the zonal direction than in the meridional direction (in which they simply oppose meridional flows). The zonal Lorentz accelerations, $\vect{B}\cdot \nabla B_x$, are most easily understood geometrically when considered as the directional derivative of $B_x$ along horizontal magnetic field lines, which are approximately equivalent to lines of constant $A$. When the magnetic field lines bend equatorward they generally move into regions of smaller $|B_x|$, hence the zonal Lorentz force component generally accelerates flows westward (as $\vect{B}\cdot \nabla B_x < 0$); conversely, when they bend poleward they generally move into regions of larger $B_x$, hence the zonal Lorentz force component generally accelerates flows eastward (as $\vect{B}\cdot \nabla B_x > 0$). One can see this by comparing lines of constant $A$ in mid-to-high latitudes of \cref{fig:Geop} (row $(b)$, middle column) with the corresponding mid-to-high latitude zonal Lorentz force accelerations in \cref{fig:dudt} (third row, middle column). Since  magnetic field lines bend equatorward between the western and eastern dayside at mid-to-high latitudes, the Lorentz force accelerates mid-to-high latitude dayside flows westward (and eastward on the nightside).

Similar westward dayside Lorentz force accelerations are generated along the equator by magnetic field lines bending into equatorial regions. To visualise this, in \cref{fig:BxEq} we plot the horizontal magnetic field geometry (top row) and the zonal component of the Lorentz force (bottom row) in the equatorial region, $-\pi/8<y/R<\pi/8$, for the transient (left) and quasi-steady (right) phases of the supercritical MHD solution (again, for parameter regime $(b)$). In the initial phase the Lorentz force primarily acts to resist drag-adjusted geostrophic circulations (see above). Therefore, in the early transient phase the magnetic field lines bend equatorward between the western and eastern dayside (where the initial circulations are poleward and equatorward respectively). {For the lowest equatorial regions ($|y/R| \lessapprox \pi/32$ in \cref{fig:BxEq}) such equatorward magnetic field line bending causes the lines to move into regions smaller $|B_x|$. Consequently, zonal Lorentz force accelerations are westward in regions surrounding the hotspot (see \cref{fig:BxEq}, lefthand column). In fact, zonal Lorentz force accelerations are always westward in hotspot regions, regardless of radiative/drag/forcing parameter choices, because in hydrodynamic (and weak/early-phase MHD) solutions hotspots are located between the substellar point and the (eastward) maximum of equatorward flow (where lines of constant $A$ are bent most equatorward). The resulting westward accelerations cause an equatorial imbalance in the zonal momentum equation (see \cref{fig:dudt}, bottom row, middle column), which drives the point of zonal equatorial wind divergence eastwards of the hotspot and, consequently, shifts the hotspot westward (see discussion in \sectref{subsect:basic:MHD}). Finally, as these westward accelerations cause dayside equatorial winds to become more westward, lines of constant $A$ are swept from east to west along the equator, bending them further and thus enhancing equatorial Lorentz force accelerations across all equatorial latitudes (see \cref{fig:BxEq}, righthand column)\footnote{This is equivalent to saying that the more westwardly-oriented  dayside winds cause $B_y$ to become more significant in equatorial regions, which in-turn enhances the westward Lorentz force accelerations.}.} 

Across radiative/drag/forcing parameter choices, when the hotspots have transitioned westwards the system rebalances into a quasi-steady state, which is characterised by westward hotspots, zonally-aligned winds, and magnetic field lines that have an equatorward bend along the line $x=0$ in equatorial regions. The predominant meridional balance is between pressure gradients, the Coriolis force, and the Lorentz force (see \cref{fig:dvdt}, righthand column); whereas the predominant zonal balance is between pressure gradients, the Lorentz force, and the Rayleigh drag (see \cref{fig:dudt}, righthand column). In these balances the zonally-aligned winds cause the meridional Rayleigh drag and the zonal Coriolis force to be small. We comment that as the magnetic field eventually diffuses away, the balance adjusts to the decreasing  Lorentz force contribution, eventually restoring the drag-adjusted geostrophic/magneto-geostrophic balances associated with hydrodynamic and weakly-magnetic solutions (and hence eastward hotspots).

\begin{figure*}
\centering
\includegraphics[scale=1]{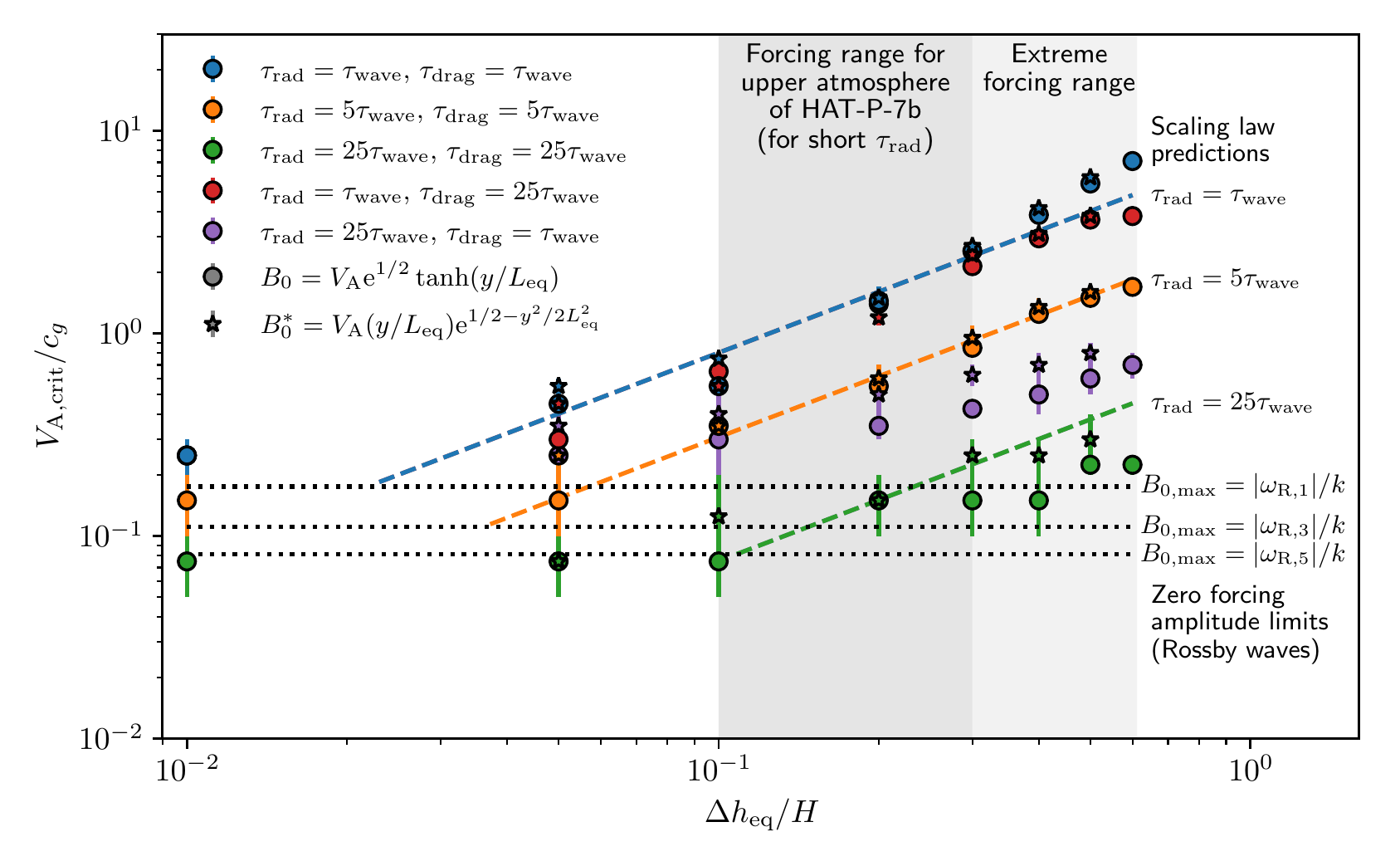}
\caption{{Quantitive dependencies of critical magnetic field amplitudes on the forcing magnitude parameter, $\Delta h_\di{eq}/H$, for different choices of $\trad$ and $\tdrag$. Critical magnetic field amplitudes are illustrated with marker points. These are located mid-way between the upper/lower bounds of the identified critical amplitude range, for a particular parameter set, with error bars indicating these upper/lower bounds. Lines indicating scaling law predictions (dashed) and  zero-amplitude limits based on the linear theory (dotted; see \sectref{sect:linear:model}) are overlaid} }  
\label{fig:VAcrit:numeric}
\end{figure*}

\subsection{Forcing dependence}  \label{subsect:Forcing:dependence}

We find that, when one compares marginally supercritical magnetic solutions with {$\Delta h_\di{eq}/H = \{0.01, 0.05$, $0.1$, $0.2$, $0.3$, $0.4$, $0.5$, $0.6 \}$}, the qualitative physical behaviours and balances discussed in \sectrefrange{subsect:basic:HD}{subsect:Force:balances} (and illustrated in \cref{fig:Geop,fig:dvdt,fig:dudt,fig:BxEq}) remain highly similar (in fact, remarkably so). The only discernible changes we observe between marginally supercritical magnetic solutions, upon increasing $\Delta h_\di{eq}/H$, are an approximately linear scaling of dependent variable magnitudes and a  correction from advection, which generally only provides a lower order correction. This is to be expected from the theory we have developed so far,  as the process that needs to be overcome in order to trigger hotspot reversals (i.e., the drag-adjusted geostrophic balance) is a linear one. Consequently, choices of $\Delta h_\di{eq}/H$ do not change the mechanics of the hotspot reversals, though they do determine quantitive features of the system (such as magnitudes and $V_\di{A, crit}$).

We can use our developed understanding of the reversal mechanism to predict magnitudes $V_\di{A, crit}$, with simple scaling arguments based on the respective magnitudes of  geostrophic circulations and the restorative Lorentz force. Let $\tau_\di{geo}^{-1}=\mathcal{U}/L_\di{eq}$ be the frequency over which geostrophic flows circulate and  $\tau_\di{A}^{-1}=\mathcal{V}/L_\di{A}$ be the (Alfv\'en) frequency over which the azimuthal field attempts to zonally-align these circulations, where $\mathcal{U}$, $\mathcal{V}$, $L_\di{eq}$ and $L_\di{A}$ are the typical velocity and length scales associated with the two opposing processes. Reversals occur when $\tau_\di{A}^{-1}\gtrsim \tau_\di{geo}^{-1}$ or equivalently when $\mathcal{V}\gtrsim \mathcal{U}L_\di{A}/L_\di{eq}$ (i.e., when the azimuthal field is strong enough to restrict the geostrophic flows). \cite{2013ApJ...776..134P} showed the velocities of geostrophic circulations in Coriolis dominated regions scale like
  \beq
 \frac{\mathcal{U}}{c_g} \sim   \left(\frac{\Delta h_\di{eq}}{H}\right)\left( \frac{\trad}{\twave}\right)^{-1}\left(\frac{2 \Omega \twave^2}{\trad} + 1 \right)^{-1},  \label{eq:U:scale:Co}
 \eeq
 highlighting that the reversal threshold is expected to have a linear dependence on $\Delta h_\di{eq}/H$.
 
In \cref{fig:VAcrit:numeric}  we plot the dependence of $V_\di{A, crit}$ on $\Delta h_\di{eq}/H$ from our simulations. For comparison, we overplot the lines
 \beq
 \begin{split}
 \frac{V_\di{A, crit}}{c_g} =  \left(\frac{  2 \pi R (\Delta h_\di{eq}/H)  }{\kappa L_\di{eq}(\trad/\twave)} \right) \hspace{-0.2em} \left(\frac{2 \Omega \twave^2}{\trad} + 1 \right)^{-1}, \label{eqn:nonlin:VAcrit}
 \end{split}
\eeq 
where, since the circulations bend field lines on the planetary azimuthal scale, we take $L_\di{A} = 2 \pi R$ and  $\kappa$ is a constant of order unity based on the profile of $B_0(y)$.\footnote{$\kappa$ is an estimate of the relative strength of $B_0$ (compared to $V_\di{A}$) at low latitudes, $y_0$, where westward Lorentz force accelerations first develop. In \cref{fig:VAcrit:numeric}, we take $\kappa =  \ee^{1/2} \tanh(y_0/L_\di{eq}) \approx 0.47$ (using $y_0\approx R \pi /16$  based on \cref{fig:BxEq}). }

 We generally find reasonable agreement between this simple scaling prediction and numerical simulations, particularly in the realistic regimes of $\trad$ short and $\Delta h_\di{eq}/H \sim 0.1$-$0.3$, but note that $V_\di{A, crit}$ approaches a minimum as $\Delta h_\di{eq}/H\rightarrow 0$, which we shall consider in \sectref{sect:linear:model}.  This scaling law approximation deals less favourably  in the (less physical) long $\trad$ cases, where $\tdrag$ dependencies become important. However, as we shall discuss in \sectsref{sect:hotspot:reversal:criterion}{sect:discussion}, the other uncertainties in atmospheric characteristics are likely to provide much larger uncertainties than those arising from this scaling law approximation.

\subsection{Linear-Gaussian magnetic field profiles} \label{subsect:B0:comparisons}
Upon comparing the discussed results to their equivalents for the initial magnetic field profile $\vect{B}(x,0)=V_\di{A}(y/L_\di{eq})\exp(1/2-y^2/2L_\di{eq}^2)$, we found the same mechanical features. Namely, subcritical solutions behave similarly to their hydrodynamic counterparts; whereas, for supercritical magnetic solutions, the obstruction of geostrophic circulations by the magnetic field causes zonal wind alignment, a westward Lorentz force acceleration, and therefore reversed hotspots. The only different qualitative flow features arise at the poles, where $V_\di{A}(y/L_\di{eq})\exp(1/2-y^2/2L_\di{eq}^2)$ decays, but our model and aims are not directed towards the polar regions. The quantitative differences between solutions are also tend to be minor, with a second order change in $V_\di{A, crit}$ as the two profiles cause a slightly different magnitude of Lorentz force to be generated for a given $V_\di{A}$. To make this comparison, we have marked $V_\di{A, crit}$ for  Linear-Gaussian profiles on  \cref{fig:VAcrit:numeric} with starred markers. We conclude that the choice of a $B_0 \propto \tanh(y/L_\di{eq})$ profile is a useful simplification when considering reversals. This can be advantageous due to properties of the hyperbolic tangent function, which is both monotonic and bounded as $y\rightarrow \infty$.

\subsection{Summary of findings} 
In this section we have identified the mechanism responsible for driving hotspot reversals in our SWMHD model. The reversals are caused by the westward Lorentz force acceleration that is generated when strong equatorially-antisymmetric azimuthal magnetic fields obstruct the  geostrophic circulation patterns responsible for energy redistribution in the hydrodynamic system. The understanding we have developed explains why such hotspot reversals always emerge in the SWMHD model, regardless of our choices for the free forcing/drag parameters $\Delta h_\di{eq}/H$, $\trad$, and $\tdrag$. Moreover, this developed understanding has allowed us to use simple scaling arguments to predict the reversal threshold, $V_\di{A, crit}$, in terms of planetary parameters, finding reasonable agreement between predictions and numerical simulations in  realistic forcing regimes for our fiducial planet HAT-P-7b. However, our simulations also show that  $V_\di{A, crit}$ approaches a minimal threshold in the zero amplitude limit. In  \sectref{sect:linear:model} we shall probe linear theory to explain this finding.  For this, we shall use our finding that, when compared, equatorially-antisymmetric azimuthal magnetic field profiles with similar latitudinal dependence at equatorial and mid-latitudes behave similarly to one another.

\section{{Linear theory}} \label{sect:linear:model} 

\subsection{{Linearised steady state solutions}} \label{sect:wave:contributions}
First we seek to establish the features of the reversals that linear theory can capture, and its limitations. We do so by linearising the non-diffusive versions of \crefrange{eqn:mom}{eqn:magfluxfunc} about the background state $\{u_0,v_0,h_0,A_0\}=\{u_0(y),0,H,A_0(y)\}$, where $H$ is the (constant) background layer thickness, $A_0$ is defined such that $\dd A_0 /\dd y = H B_0$ for the latitudinally-dependent azimuthal background magnetic field, $\vect{B}_0=B_0(y)\uvect{x}$, and $u_0(y)$ is to be fixed in a manner that balances the zeroth order zonal momentum equation of the hydrodynamic version of the system which we wish to investigate. To probe the system at the reversal threshold, we assume steady state perturbations exist about this background state and apply the plane wave ansatz, $\{u_1,v_1,h_1,A_1\}=\{\hat{u}(y),\hat{v}(y),\hat{h}(y),\hat{A}(y)\} \e^{ikx}$, where $k$ denotes the azimuthal wavenumber and subscripts of unity denote perturbations from the background state. Such perturbations satisfy
\beq
\begin{split}
    (i k u_0 +  \tdrag^{-1}) \hat{u}   = &  \left( f - \oderiv{u_0}{y} \right) \hat{v}  - ikg \hat{h}  \\ 
    & + ikB_0 \hat{B}_x + \oderiv{B_0}{y} \hat{B}_y,  
\end{split} \label{eq:SS:mom:x} 
\eeq
\begin{align}
    (i k u_0 +  \tdrag^{-1})   \hat{v}  = &  -f \hat{u}  - g \oderiv{\hat{h}}{y} + ikB_0 \hat{B}_y, \label{eq:SS:mom:y} \\ 
    (i k u_0 +  \trad^{-1}) \hat{h}    =& - H \left(  ik\hat{u} + \oderiv{\hat{v}}{y} \right) + H S(y), \label{eq:SS:cont}  \\ 
   i k u_0  \hat{A}  =& -H B_0 \hat{v}, \label{eq:SS:ind}
\end{align}  
where $\hat{B}_x = (\textoderiv{\hat{A}}{y}-B_0 \hat{h})/H$, $\hat{B}_y = -ik\hat{A}/H$, $S(y)=(\Delta h_\di{eq}/H)\,\trad^{-1}\exp(-y^2/2L_\di{eq}^2)$ is the first order forcing contribution in the system based on the equilibrium thickness profile,  $h_\di{eq} =  H + \Delta h_\di{eq} \cos \left( k x \right) \exp(-y^2/2L_\di{eq}^2)$, and based on our numerical findings we have assumed that $\vect{R}$ does not make a first order contribution to \cref{eq:SS:mom:x,eq:SS:mom:y}. Before solving, we note that hydrodynamic solutions are never singular, but  that \cref{eq:SS:ind} causes the magnetic version of the system to be singular if $u_0=0$.  To compare to the simulations of \sectref{sect:numerical:results}, we solve the system for $f=\beta y$ and $B_0 = V_\di{A} \ee^{1/2} \tanh(y/L_\di{eq})$. 

 \begin{figure*}
\centering
\includegraphics[scale=1]{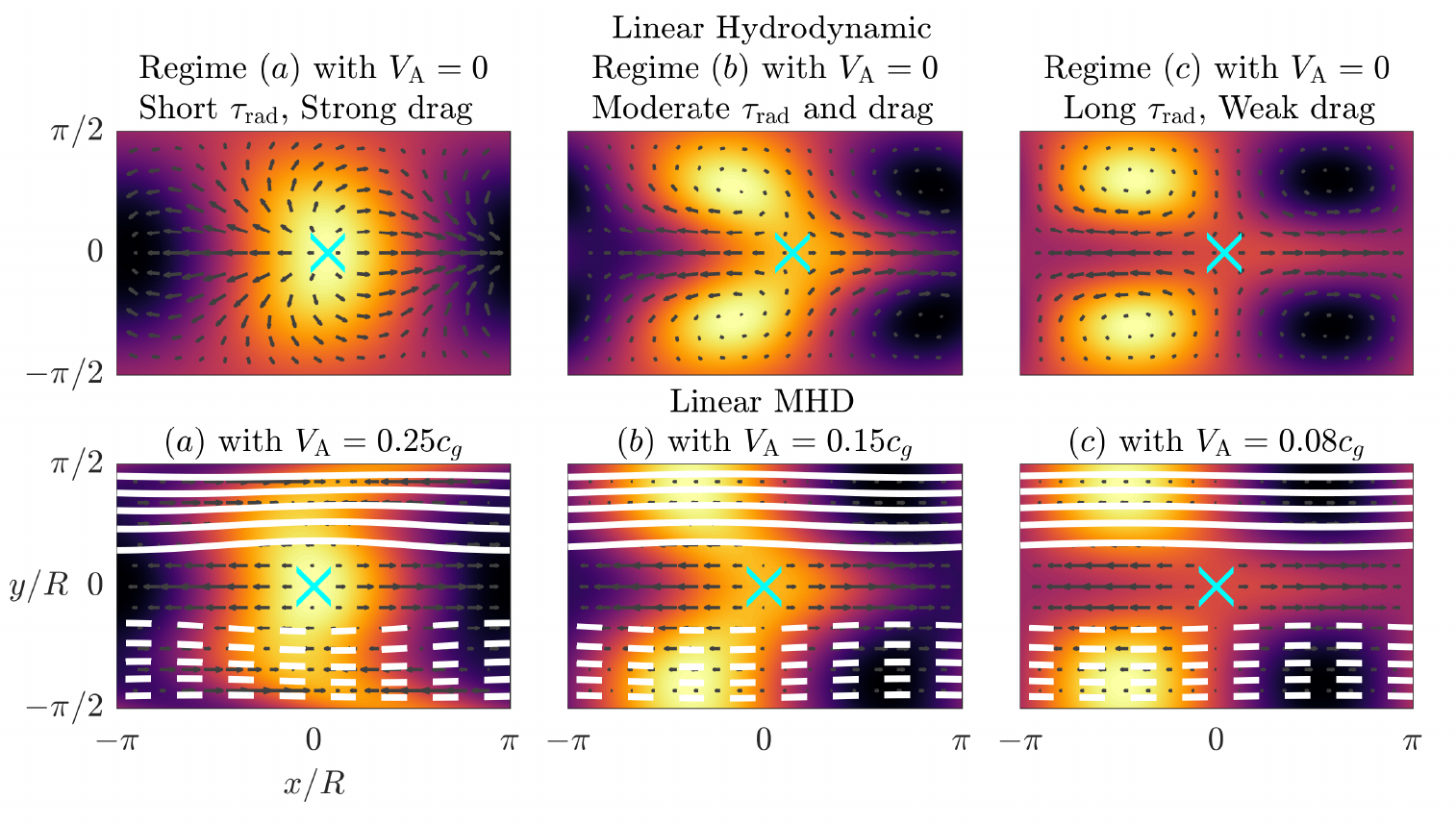}  
 \caption{Linear solutions (for $\Delta h_\di{eq}/H = 0.01$ and $k=1/R$). Contours of the relative layer thickness deviations (rescaled geopotential energy deviations) are plotted on (individually-normalised) colour axes, with (individually-normalised) velocity vectors ($\vect{u}=\vect{u}_0+\vect{u}_1$), hotspots (cyan crosses) and, where relevant, lines of constant $A=A_0+A_1$ (solid/dashed for $B_x$ positive/negative) over-plotted. Hydrodynamic solutions (top row) are compared to marginally critical MHD solutions (bottom row; compare $V_\di{A}$ values to \cref{fig:VAcrit:numeric}). Solutions are plotted for $(a)$ $\trad = \tdrag = \twave$ (left); $(b)$ $\trad = \tdrag = 5\twave$ (middle);  and  $(c)$ $\trad = \tdrag = 25 \twave$ (right). Solutions are calculated for $-5L_\di{eq}<y<5L_\di{eq}$, but  are cut off for $-R \pi/2<y<R \pi/2$ (recall, $L_\di{eq}/R\approx0.67$). The strong magnetic field aligns flows preventing geopotential recirculation between latitudes, but in the linearised model the (non-linear) equatorial Lorentz force is zero. Consequently, in the linearised model hotspot offsets of marginally critical MHD solutions tend to zero, but do not reverse like full SWMHD simulations.} \label{fig:steady:linear:solutions} 
\end{figure*}

For a given $u_0(y)$, we seek solutions of \crefrange{eq:SS:mom:x}{eq:SS:ind} on $-L_y<y<L_y$, with impermeable boundaries at $y=\pm L_y$, using the shooting method outlined in  \appref{app:forced:shooting}. We take $L_y=5L_\di{eq}$ (see \cref{eqn:Leq}), which is large enough to ensure that the outer boundary condition has a negligible influence on solutions. We solve the system for $u_0(y)=U_0 \exp(-y^2/2L_\di{eq}^2)$, where $U_0$ is chosen so that in the hydrodynamic limit the zonally-averaged zonal-acceleration in Equation (22) of \cite{2011ApJ...738...71S} vanishes at the equator.  We plot linear solutions for $\Delta h_\di{eq}/H = 0.01$ in \cref{fig:steady:linear:solutions}, on the reduced domain $-\pi/2<y/R<\pi/2$, for three $\trad$ and $\tdrag$ choices, comparing hydrodynamic solutions with MHD solutions at the threshold of criticality, as found by simulations.

Hydrodynamic solutions generally resemble those discussed in \cite{2011ApJ...738...71S}, albeit with an adjustment due to $u_0$ \cite[as discussed by][for $u_0$ constant]{2014ApJ...793..141T}. They are characterised by geostrophic circulations at mid-to-high latitudes and zonal pressure driven jets at equatorial latitudes. Such solutions closely resemble the non-linear hydrodynamic steady state solutions we discussed in \sectref{sect:numerical:results}. The characteristic flow patterns of hydrodynamic steady state solutions can also be directly linked to the forcing responses of specific standing, planetary scale, equatorial shallow-water waves \citep{196625,2011ApJ...738...71S,2014ApJ...793..141T}. The geostrophic circulations are linked to the planetary scale equatorial Rossby waves, which are geostrophic in nature at mid-to-high latitudes; while the equatorial jets are linked the superposition of the planetary scale equatorial Rossby waves and the equatorial Kelvin wave, which travels eastward about the equator in response to pressure perturbations. The presented linear hydrodynamic solutions all have eastward hotspots (located at points of zonal wind divergence), as the linearised meridional convergence of geopotential flux into the equator, $-gH \textderiv{v_1}{y}|_{y=0}$, is maximised eastward of the substellar point (due to the form of the geostrophic circulations; further discussion in \sectref{subsect:basic:HD}).

The marginally critical MHD solutions share some common characteristics with their non-linear simulated counterparts. Specifically, in these solutions the aligning influence of the meridional Lorentz force is strong enough obstruct geostrophic circulations, which are replaced by zonally-aligned winds. However, unlike their simulated non-linear counterparts, the magnetohydrodynamic solutions do not have westward hotspots. The arises because in this simple linear model one can show that the Lorentz force components, which drive hotspots reversals in non-linear simulations (see  \sectref{sect:numerical:results}), vanish at the equator.\footnote{For $S(y)$ equatorially-symmetric and $B_0$ equatorially-antisymmetric,  $\hat{v}$ is antisymmetric and $\{\hat{u},\hat{v},\hat{A}\}$ are symmetric about the equator (see \appref{app:forced:shooting}).   Hence, $\hat{B}_x=(\textoderiv{\hat{A}}{y}-B_0 \hat{h})/H$ is antisymmetric; while, by  \cref{eq:SS:ind}, $\hat{A}(0)=0$, so $\hat{B}_y(0)=-ik\hat{A}(0)/H=0$. Consequently,  $ikB_0 \hat{B}_x + \textoderiv{B_0}{y} \hat{B}_y$ and $ikB_0 \hat{B}_y$ both vanish at the equator. } Instead, marginally critical MHD solutions approach a limit of zero hotspot offset, as the obstruction of geostrophic circulations causes $-gH \textderiv{v_1}{y}|_{y=0}\rightarrow 0$. This highlights that in simple linear models, with similar linearisations of the Lorentz force and the induction equation (i.e., without more sophisticated treatments of magnetic diffusion and non-linear effects), one can identify the obstruction of geostrophic circulations that cause hotspot reversals in non-linear simulations, but not westward hotspot offsets explicitly. This observation is useful in the remainder of this section, where we aim to link the magnetic obstruction of geostrophic circulation patterns to wave dynamics.

 \subsection{{Wave background: Alfv\'en-Rossby wave coupling}} 
 
Various authors have studied the linear waves present in rotating MHD systems. Early studies, which used quite general (usually uniform) flow/field geometries, focussed on the influence that these waves have on the geodynamo \citep{1966RSPTA.259..615H,1969JFM....39..283H,Acheson_1973}. Since the development of SWMHD \citep{2000ApJ...544L..79G}, authors have been able to utilize its reduced geometry to study waves in more specific flow/field geometries. Rotating SWMHD waves have now been studied for a variety of thin-layered astrophysical and geophysical systems including the geodynamo, the solar tachocline, and neutron star atmospheres \citep{2001ApJ...551L.185S,2007A&A...470..815Z,2009ApJ...691L..41Z,2009ApJ...703.1819H,2017GApFD.111..282M,2018ApJ...856...32Z}. Of these, the solar tachocline, which is also expected to have an equatorially-antisymmetric toroidal dominant magnetic field geometry, can be considered as 
 similar to the hot Jupiter system. \cite{2001ApJ...551L.185S} studied waves in the local regions of the solar tachocline, focusing on regions away from the equator; whereas  \cite{2007A&A...470..815Z,2009ApJ...691L..41Z} studied the global dynamics of these waves for the two extreme cases ($\epsilon \gg 1$ and $\epsilon \ll 1$) of the rotation-stratification parameter, $\epsilon = 4 \Omega^2 R^2 / c_g^2$. However, the atmosphere of HAT-P-7b lies in the region of parameter space between these two extremes ($\epsilon \approx 4.8$). \cite{2018ApJ...856...32Z} studied equatorial SWMHD waves using an equatorial beta-plane model for two purely-azimuthal magnetic field geometries: firstly, uniform and, secondly, equatorially-antisymmetric (latitudinally-linear). \cite{2017GApFD.111..115L} and \cite{2018GApFD.112...62L} studied some asymptotic solutions of the beta-plane and spherical version of the system, with an equatorially-antisymmetric  azimuthal field, in certain weak and strong field limits, but we wish to study the transition where magnetism becomes dynamically important. The predictions made in \cite{2019ApJ...872L..27H} were based on the equatorially-antisymmetric azimuthal magnetic field results of \cite{2018ApJ...856...32Z}. However, in this section we relax the weakly-magnetic assumptions that those analyses take. 
 
 Past works of linear waves in rotating MHD systems find that Alfv\'en waves and Rossby waves are coupled. To illustrate this, we highlight the known local dispersion relations of waves in the non-diffusive, unforced, drag-free SWMHD system, with the uniform azimuthal background magnetic field, $\vect{B}_0=V_{\di{A},0} \uvect{x}$, and the generalised beta-plane treatment $f=f_0+\beta y$.\footnote{Here $f_0 \equiv f(y_0) = 2\Omega \sin \theta_0$ and  $\beta \equiv \textoderiv{f}{y}|_{y=y_0} = (2 \Omega/R) \cos \theta_0$ are respectively the Coriolis parameter and the Coriolis parameter's local latitudinal variation at a reference latitude, $\theta_0 = y_0/R$, about which the system is centred. }  In local regions (i.e., if  $|y/R|\ll1$ and $|\beta y| \ll |f_0|$), this linearised system associated with this background state may be approximately solved with the plane wave ansatz: $\{u_1,v_1,h_1,A_1\}=\{\hat{u},\hat{v},\hat{h},\hat{A}\} \e^{i(kx+ ly -\omega t)}$, where hatted variables are constant amplitudes of the plane wave solutions, $k$ is the azimuthal wavenumber, $l$ is the latitudinal wavenumber, and $\omega$ is the oscillation frequency. Seeking solutions that are first order in the Coriolis parameter only, yields the following dispersion relation \citep{2007A&A...470..815Z,2009ApJ...703.1819H}:
\beq
\begin{split}
\omega^4 - \omega^2(K^2 c_g^2 + 2 k^2 V_{\di{A},0}^2 + f_0^2)  - \omega \beta k c_g^2 & \\ + k^2 V_{\di{A},0}^2 (K^2 c_g^2 + k^2 V_{\di{A},0}^2)& = 0, \label{eqn:dispersion:all:C3}
\end{split}
\eeq
for arbitrary wave amplitudes, where $K \equiv (k^2 + l^2)^{1/2}$. For $V_{\di{A},0}>0$, \cref{eqn:dispersion:all:C3} has four solutions, which in the rotation-free limit ($f_0 = \beta = 0$) are \citep{2001ApJ...551L.185S}
\beq
\omega^2 = 
\begin{cases}
    V_{\di{A},0}^2 k^2, \\  
    c_{m,0}^2 k^2 +  c_g^2 l^2, 
\end{cases}  \label{eqn:alfven:magnetogravity:waves} 
\eeq
where $c_{m,0} = (c_g^2 + V_{\di{A},0}^2)^{1/2}$ is the magneto-gravity wave speed for a constant background magnetic field. The first pair of solutions are Alfv\'en waves, which are driven by magnetic tension and travel parallel to the background magnetic field; the second pair of solutions are magneto-gravity waves, which propagate horizontally to restore pressure gradients and magnetic tension. For rotationally modified waves, we follow \cite{2001ApJ...551L.185S} by labelling the rotationally modified Alfv\'en waves as slow ``Alfv\'en branch'' solutions and the rotationally modified magneto-gravity waves as fast ``magneto-gravity branch'' solutions.
 
In the fast wave limit ($|\omega|/|2\Omega|\gg 1$), to leading order, fast magneto-gravity branch solutions satisfy \cite[][but with $\vect{B}_0=V_{\di{A},0} \uvect{x}$]{2009ApJ...703.1819H}
\beq
\begin{split}
\omega^2 & \approx \frac{K^2 c_g^2}{2} + k^2 V_{\di{A},0}^2 + \frac{f_0^2}{2} \\ & + \frac{1}{2} \sqrt{K^2 c_g^2( K^2 c_g^2 + 2f_0^2 ) + f_0^2 ( f_0^2 + 4 k^2 V_{\di{A},0}^2 )},\label{eqn:solution:poincare}
\end{split}
\eeq
These two magneto-gravity branch solutions travel in opposite directions in order to restore pressure gradients and magnetic tension, but with a Coriolis modification. Solutions of this kind are known as magneto-Poincar\'e waves \citep[as they reduce to Poincar\'e waves for $V_{\di{A},0}=0$; e.g., ][]{2009ApJ...703.1819H} or magneto-inertial gravity waves  \cite[and inertial gravity waves in hydrodynamics; e.g.,][]{2017GApFD.111..282M,2018ApJ...856...32Z}. We choose the inertial gravity and magneto-inertial gravity nomenclature (IG and MIG hereafter). The independence of \cref{eqn:solution:poincare} on $\beta$ highlights that IG/MIG solutions do not generally have a leading order  dependence on $\beta$ \citep[and exist on the f-plane; e.g.,][]{2006aofd.book.....V}. 

In the slow wave ($|\omega|/|2 \Omega | \ll 1$) limit, the dispersion relation evaluated at the equator ($f_0=0$ and $\beta = 2\Omega / R$), yields the two Alfv\'en branch solutions \cite[][but with $\vect{B}_0=V_{\di{A},0} \uvect{x}$]{2009ApJ...703.1819H}:
\beq
    \omega = \frac{-\beta k c_g^2 \mp \sqrt{(\beta k c_g^2)^2 + M } }{2 ((c_{m,0}^2 + V_{\di{A},0}^2) k^2 +  c_g^2 l^2) },  \\
    \label{eqn:Rossby:Alfven:coupling}
\eeq
where $M = 4 k^2 V_{\di{A},0}^2 (c_{m,0}^2 k^2 +  c_g^2 l^2)((c_{m,0}^2 + V_{\di{A},0}^2) k^2 +  c_g^2 l^2)$ is the magnetic component of the numerator.  In the limit where these solutions are dominated by the Alfv\'en speed, these waves  Alfv\'enic in nature (see by taking $V_{\di{A},0}$ dominatingly large).  Conversely, in the hydrodynamic limit ($V_{\di{A},0} = 0$) the Alfv\'en branch solutions reduce to
\beq
{\omega =}
\begin{cases}
    -\beta k / (k^2+l^2), \\
    0, 						
\end{cases}  \label{eqn:rossby:wave:solution}
\eeq
so the eastward Alfv\'en branch solution vanishes and the westward Alfv\'en branch solution reduces to a Rossby wave.

 This Alfv\'en-Rossby wave coupling is a well-documented feature of MHD in systems with a latitudinally dependent planetary vorticity \citep{1966RSPTA.259..615H,1969JFM....39..283H,Acheson_1973}. However, Alfv\'en and Rossby waves are fundamentally different in nature. Rossby waves arise due to potential vorticity conservation and the latitudinal variation of the Coriolis parameter. They behave geostrophically and are highly dispersive, so can transfer energy and angular momentum to the surrounding system \cite[e.g.,][]{2006aofd.book.....V,pedlosky2013waves}. Conversely, pure Alfv\'en waves travel parallel to the dominant azimuthal magnetic field geometry and are non-dispersive, so travel as zonally-aligned solitons. Comparing the oscillation frequency of Rossby ($\omega_\di{R}$) and Alfv\'en ($\omega_\di{A}$) waves gives
 \beq
 |\omega_\di{R}/\omega_\di{A}| = \beta / V_{\di{A},0} (k^2+l^2),
 \eeq
 suggesting that, for given choices of $\beta$ and $V_{\di{A},0}$, Rossby wave characteristics dominate at large scales; whereas Alfv\'en wave characteristics dominate at small scales. In \sectref{sect:numerical:results}, we showed that reversals on hot Jupiters are closely tied to the zonal-alignment of equatorially-adjacent geostrophic circulations by equatorially-antisymmetric azimuthal magnetic fields. Therefore, to investigate reversals in the zero amplitude limit,  we examine the behaviour of equatorial waves as the \Alf oscillation frequency approaches $\omega_\di{R}$  in magnitude for an antisymmetric azimuthal background magnetic field.

\subsection{{Equatorial magnetohydrodynamic wave equations}} \label{subsect:linear:model}
To study the linear equatorial magnetohydrodynamic waves of the system, we linearise the non-diffusive, unforced, drag-free versions of \crefrange{eqn:mom}{eqn:magfluxfunc}  about the background state, $\{u_0,v_0,h_0,A_0\}=\{0,0,H,A_0(y)\}$, where $H$ is the constant and $\dd A_0 /\dd y = H B_0$ (for $\vect{B}_0=B_0(y)\uvect{x}$ in velocity units). Applying the plane wave ansatz, $\{u_1,v_1,h_1,A_1\}=\{\hat{u}(y),\hat{v}(y),\hat{h}(y),\hat{A}(y)\} \e^{i(kx-\omega t)}$, the evolution of the perturbations is determined by the following linearised SWMHD system:
\begin{align}
    &-i\omega \hat{u}   = f \hat{v}  - ikg \hat{h} + ikB_0 \hat{B}_x + \oderiv{B_0}{y} \hat{B}_y,  \label{eqn:mom:lin1} \\ 
    &-i\omega \hat{v}   = -f \hat{u}  - g \oderiv{\hat{h}}{y} + ikB_0 \hat{B}_y, \label{eqn:mom:lin2} \\ 
    &-i\omega \hat{h}    = - H \left(  ik\hat{u} + \oderiv{\hat{v}}{y} \right),  \label{eqn:cont:lin}  \\ 
    &-i\omega \hat{A} = -H B_0 \hat{v},  \label{eqn:ind:lin}  
\end{align}  
where $\hat{B}_x = (\textoderiv{\hat{A}}{y}-B_0 \hat{h})/H$ and $\hat{B}_y = -ik\hat{A}/H$. From this we eliminate $\hat{u}$, $\hat{h}$, $\hat{A}$, $\hat{B}_x$, and $\hat{B}_y$ to obtain the single ordinary differential equation:
\beq
\mathcal{L}\{ \hat{v} \} \equiv F_1 \osecderiv{\hat{v}}{y} + F_2 \oderiv{\hat{v}}{y} + F_3 \hat{v} = 0, \label{eqn:wave:ODE}
\eeq
for the latitudinal solving domain, $-L_y < y < L_y$, with
\begin{fleqn}[1em]
\begin{eqnarray}
 & F_1 =&  \left(\omega^2-B_0^2k^2 \right)\left(\omega^2-c_m^2k^2\right), \label{eqn:ODE:term1:eval} \\
 & F_2  =& 2B_0\oderiv{B_0}{y}c_g^2 k^4,  \label{eqn:ODE:term2:eval} \\
& \begin{split} F_3 = & \frac{(\omega^2-c_m^2k^2)}{c_g^2} \left[ (\omega^2-c_m^2k^2)(\omega^2-B_0^2k^2) \phantom{\oderiv{f}{y}}\right. \\
&\quad \left. - \omega^2f^2 - \omega k \oderiv{f}{y} c_g^2 \right] - 2\omega f B_0 \oderiv{B_0}{y} k^3,
\end{split} \label{eqn:ODE:term3:eval}
\end{eqnarray}
\end{fleqn}
where $c_m(y) \equiv (c_g^2+B_0^2)^{1/2}$ denotes the (rotationless) magneto-gravity wave speed. This system can the contain singular points at $y=y_s$, if $\omega = \pm B_0(y_s) k$ (Alfv\'en singularity) or $\omega = \pm c_m(y_s) k$ (magneto-gravity singularity), which we label based on the $\omega$-regions each singularity is associated with. 

If one attempts to write $\mathcal{L}$ in Sturm-Liouville form\footnote{We use the Sturm-Liouville definition: $(p \hat{v}')' + q \hat{v} = \lambda w  \hat{v}$, where $p(y), w(y) > 0$, and $p(y)$, $p'(y)$, $q(y)$, and $w(y)$ are continuous functions over the system's finite solving domain, $y\in [-L_y,L_y]$.}, through use of an integrating factor, it is found that the highest order functional coefficient of the Sturm-Liouville operator, $p=(\omega^2-B_0^2k^2)/ (\omega^2-c_m^2k^2)$, is not independent of the oscillation frequency. Therefore, the desirable properties of the Sturm-Liouville eigenvalue problem (e.g.,~real eigenvalues and orthogonality of eigenfunctions) are not generally guaranteed. \cite{2018ApJ...856...32Z}  studied this system in the weakly-magnetic limit where singular points do not influence the planetary scale waves.\footnote{Precisely,  \cite{2018ApJ...856...32Z} used $B_0=\gamma y$ with constant $\gamma$, applying the weak-field assumptions $\omega^2 \gg  \gamma^2 k^2 y^2$ and $|\omega^2 - c^2 k^2 | \gg  \gamma^2k^2 y^2$.}
In this approximation $\mathcal{L}$ can be re-expressed in terms of the parabolic cylinder Sturm-Liouville operator \citep[the hydrodynamic version of $\mathcal{L}$, see ][]{196625}. Therefore, away from singular $\omega$-regions, where the approximations of \cite{2018ApJ...856...32Z} hold, one may expect solutions to conform to Sturm-Liouville properties (which we find in the following analysis).

\subsection{Equatorial wave solving method} \label{subsect:lin:wave:method} 
We now examine non-trivial eigenvalue-eigenfunction pairs, $\{\omega,\hat{v}(y)\}$, that satisfy $\mathcal{L}\{\hat{v}\}=0$ everywhere in the latitudinal domain, $-L_y < y < L_y$, subject to impermeable boundary conditions (i.e., $\hat{v}(\pm L_y)=0$). We use the planetary parameters discussed in \sectref{sect:numeric:model}, $f=\beta y$ and $B_0=V_\di{A}\ee^{1/2}\tanh(y/L_\di{eq})$. {This $B_0(y)$ choice is useful because it is both monotonic and bounded as $y\rightarrow \infty$   \citep{2017GApFD.111..115L}, so there is at most one \Alf singularity in each hemisphere. For this $B_0(y)$ choice, solutions with $c_g k \leq |\omega| \leq (c_g^2 + V_\di{A}^2 \ee )^{1/2}k$ have magneto-gravity singularities; while solutions with $ |\omega| \leq V_\di{A} \ee^{1/2} k$ have Alfv\'en singularities.} We seek wave-like solutions with the planetary scale azimuthal wavenumber, $k=1/R$. We find that solving this eigenvalue problem, without further approximation on $\mathcal{L}$, is an analytically intractable problem so we use a semi-analytic approach.

Since $\mathcal{L}$ is symmetric about the equator, homogeneous solutions will be either {\em symmetric} ($\hat{v}$ symmetric and $\hat{u}, \hat{h}, \hat{A}$ antisymmetric) or {\em antisymmetric} ($\hat{v}$ antisymmetric and $\hat{u}, \hat{h}, \hat{A}$ symmetric) about the equator.\footnote{If $\hat{v}$ is equatorially-symmetric, \crefrange{eqn:mom:lin1}{eqn:ind:lin} yield $\hat{u}, \hat{h}, \hat{A}$ antisymmetric and vice versa.}  Although the system we solve here is unforced, we wish to compare solutions to the numerical simulations of \sectref{sect:numerical:results}, which had equatorially-symmetric forcing on $h$. Therefore, we only consider antisymmetric homogeneous solutions and solve $\mathcal{L}\{\hat{v}\}=0$ in the upper-half domain, $0<y<L_y$, with the antisymmetric lower boundary condition $\hat{v}(0) = 0$, which replaces $\hat{v}(-L_y)=0$. Eigenfunctions are defined up to a constant factor, so a third and final normalisation boundary condition must also be included. We set $\dd \hat{v} / \dd y |_{y=0} = \mathcal{N}$, where $\mathcal{N}$ is a normalisation constant chosen for numerical convenience, and take $L_y = 5 L_\di{eq}$  to ensure boundary influences are negligible. 

We use a shooting method to seek eigensolutions. The shooting method calculates successive ``shots'' (or test solutions, $\hat{v}_T$) for given test frequencies, $\omega_T$, where each shot satisfies $\mathcal{L}\{\hat{v}_T\}=0$, subject to two of the three boundary conditions. The third boundary condition is then satisfied by varying $\omega_T$ so that the deviation from the 
third boundary condition, $G[\omega_T]$, vanishes. 

If the system has no singular points, shots are carried out by the inversion of the tridiagonal matrix that corresponds to \eqnref{eqn:wave:ODE}, with finite difference discretizations, such that the lower boundary conditions are satisfied. We find that magneto-gravity singularities are false singularities (i.e., $\mathcal{L}$ is singular but solutions are not; see \appref{app:sing:solns}), so, for $V_\di{A}>0$, solutions in the magneto-gravity singularity $\omega$-range can also be treated as regular everywhere. For solutions in the Alfv\'en singularity $\omega$-range, we construct Frobenius power series solutions in the singular region (see \appref{app:sing:solns}), fix constants of integration by shooting into, and matching with, the $y=0$ boundary conditions, before finally shooting towards $y=L_y$ to obtain $G[\omega_T]$. Solutions are then checked via back-substitution. 

As discussed above, Sturm-Liouville theory only guarantees real eigenvalues in the weakly-magnetic limit. Therefore, we examine convergence for complex test frequencies, which have $G= G_r + i G_i = 0$ for $G_r, G_i \in \mathbb{R}$. We find that $G_r G_i$ is antisymmetric about $\omega_i = 0$, with contours $G_r=0$ and $G_i=0$ crossing exclusively on the real line, so $\omega \in \mathbb{R}$. We find the position of eigensolutions on the real line using the bracketed Newton-Raphson method discussed in \cite{1992nrfa.book.....P}. 

\subsection{Free wave eigensolutions} \label{subsect:eigensolutions}
\begin{figure}
\centering
\includegraphics[scale=1]{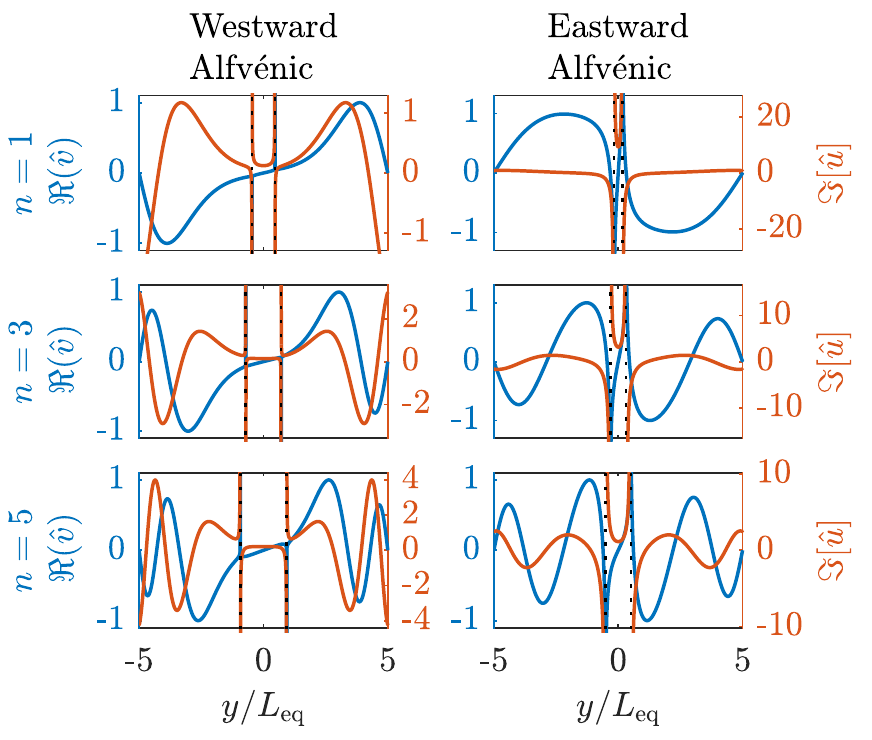}
\caption{The velocity profiles of the first few singular free wave eigenfunctions are plotted for  $V_\di{A} = 0.2 c_g/R$ and $k=1/R$. Magnetic systems have two sets of singular solutions: one westward travelling and one eastward travelling, which have Alfv\'enic properties (see main text).  $\hat{v}$ (blue) and $\hat{u}$ (red) are respectively purely real and purely imaginary for the normalisation we apply. We mark asymptotes at $y=\pm y_s$ with dotted black lines. The solutions are labelled with the latitudinal mode number, $n$, based on the latitudinal dependence of $\hat{v}$.  The corresponding profiles for $V_\di{A} = 0.15 c_g/R$ are qualitatively identical. }
\label{fig:Alf:solutions}
\end{figure}

\begin{figure*}
\centering
\includegraphics[scale=1]{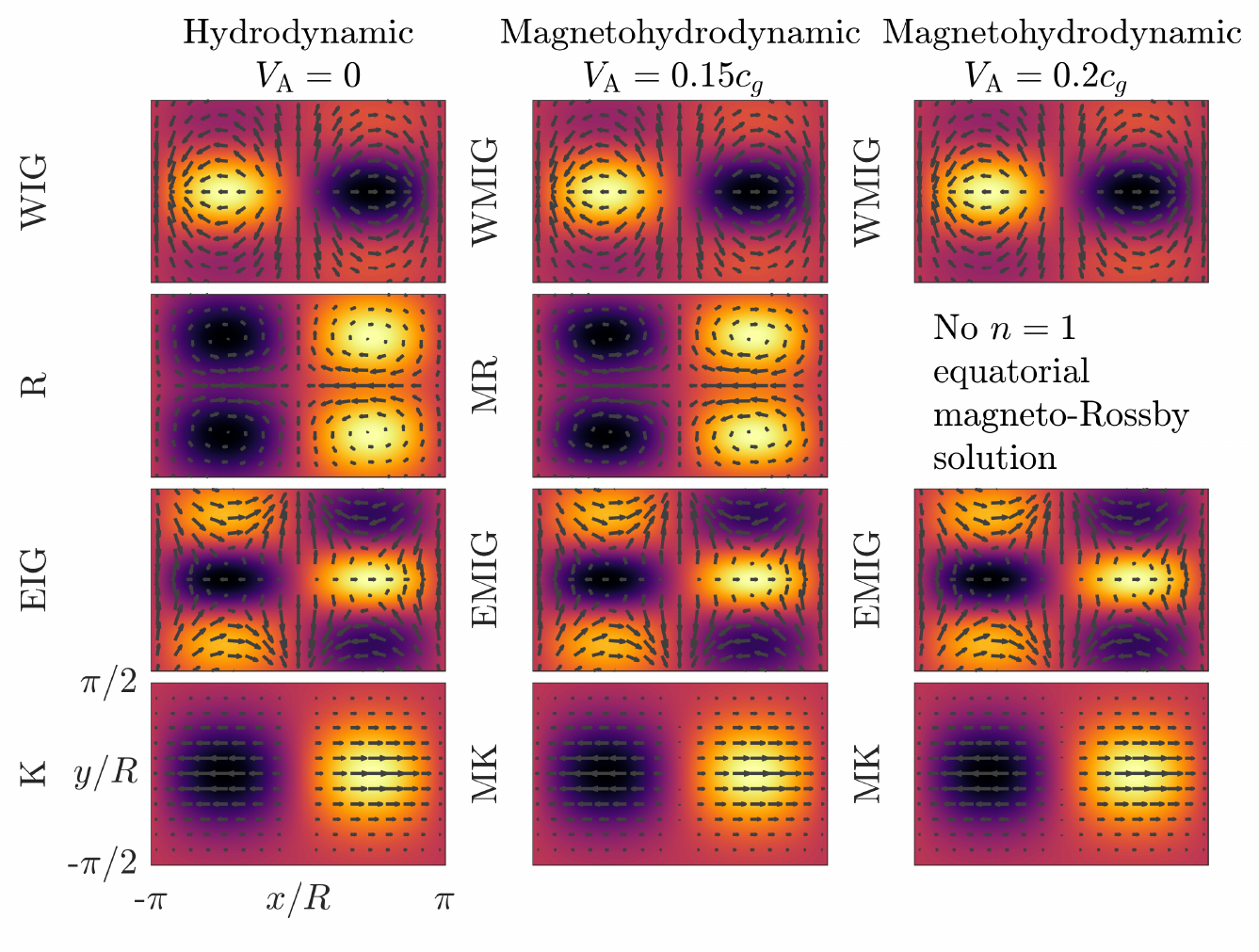}
\caption{The regular equatorial $n=1$ (rows one to three) and $n=-1$ (row four) free wave eigenfunctions (geopotential contours with overlaid velocity vectors) are plotted for $V_\di{A}=0$, $V_\di{A}=0.15c_g$, and $V_\di{A}=0.2c_g$, taking $k=1/R$. We label rows according to their wave types (see \cref{tab:eigenvalues}). Solutions are calculated for $-5L_\di{eq}<y<5L_\di{eq}$, but  are cut off for $-R \pi/2<y<R \pi/2$ ($L_\di{eq}/R\approx0.67$).} \label{fig:eigenfunctions}
\end{figure*}

We label non-singular eigensolutions with a meridional mode number, $n$, based on the hydrodynamic convention. Generally, when the domain is finite and large enough, magnetic eigenfunctions for solutions without singularities are qualitatively similar to their hydrodynamic counterparts and $n$  is the number of internal points where $\hat{v}(y)=0$ in $-L_y < y< L_y$. However, hydrodynamic Kelvin solutions have the property $\hat{v}=0$ everywhere so represent a special case. They are typically labelled with the meridional mode number $n=-1$, with $\psi_{-1}=0$ \citep{196625}. We find that solutions with $c_g k \leq |\omega| \leq (c_g^2 + V_\di{A}^2 \ee)^{1/2} k $ are the magnetic versions of Kelvin solutions, so we label them with $n=-1$ for consistency, although we find they have small non-zero $\hat{v}$ (see below). For hydrodynamic and weakly-magnetic systems there are three solutions for each $n \geq 1$: one equatorial Rossby/magneto-Rossby solution, one westward equatorial IG/MIG solution, and one eastward equatorial IG/MIG solution. When magnetism is included another two sets of solutions (one east; one west), with $ |\omega| \leq V_\di{A} \ee^{1/2} k$, emerge. These solutions, which have \Alf singularities (where $\omega^2 = B_0(y_s)^2 k^2$), differ significantly from regular equatorial wave solutions (see below). For convenience, we label these with a meridional mode number, $n$, determined by the scale of latitudinal variations in $\hat{v}$ (for $n=1,3,5$, $\hat{v}$ is plotted in \cref{fig:Alf:solutions}). In \cref{tab:eigenvalues} we present oscillation frequencies, $\omega$, for the  $n=1$, $n=3$, and $n=-1$ free wave eigensolutions, with each row representing a specific type of equatorial wave (see caption).  We present the oscillation frequencies for $V_\di{A} = 0$, $V_\di{A} = 0.15 c_g/R$ and  $V_\di{A} = 0.2 c_g/R$ and, in cases where eigenfunctions are finite everywhere, we plot the corresponding free wave eigenfunctions for the equatorial  $n=1$ and $n=-1$ waves in \cref{fig:eigenfunctions}. 

\begin{deluxetable}{Cc|C|C|C}
\tablecaption{Oscillation frequencies, $\omega$, for the $n=1$, $n=3$, and $n=-1$ equatorial wave solutions with the planetary scale azimuthal wavenumber, $k=1/R$, are tabulated for three choices of $V_\di{A}$. In the {\em Solution type} column we use the following shorthands: R/MR denotes Rossby/magneto-Rossby solutions, WIG/WMIG denotes westward  inertial gravity/magneto-inertial gravity solutions, EIG/EMIG denotes eastward inertial gravity/magneto-inertial gravity solutions, WA denotes (singular) westward Alfv\'en solutions, EA denotes (singular) eastward Alfv\'en solutions, K/MK denotes equatorial Kelvin/magneto-Kelvin solutions, and BK/BMK denotes boundary Kelvin/magneto-Kelvin solutions. \label{tab:eigenvalues}}
\tablehead{
\multicolumn2c{}  & 
\colhead{$V_\di{A} = 0$}  & 
\colhead{$V_\di{A} = 0.15 c_g$} &
\colhead{$V_\di{A} = 0.2 c_g$}  \\
\colhead{$n$} & 
\colhead{Solution type} & 
\colhead{$\omega /(c_g/R)$}  & 
\colhead{$\omega /(c_g/R)$}  & 
\colhead{$\omega /(c_g/R)$}  
}   
\startdata
1 & WIG/WMIG & -2.57   & -2.61   & -2.62 \\
1 & R/MR          & -0.293 & -0.326 & {*}     \\    
1 & EIG/EMIG   & 2.89    & 2.90    &  2.91 \\      
1 & WA$^\dagger$  & {*} & -0.117$^\dagger$  &  -0.142^\dagger \\    
1 & EA$^\dagger$   & {*} & 0.0329$^\dagger$ &  0.0556^\dagger \\   
3 & WIG/WMIG & -3.98   & -3.99   & -4.00 \\
3 & R/MR          & -0.134 & {*}       & {*}     \\    
3 & EIG/EMIG   & 4.11    & 4.12    &  4.13  \\      
3 & WA$^\dagger$  & {*} & -0.161$^\dagger$  &  -0.201^\dagger \\    
3 & EA$^\dagger$   & {*} & 0.0640$^\dagger$ &  0.102^\dagger \\    
-1 & K/MK          & 1         & 1.01 & 1.01    \\     
-1 & BK/BMK          & -1         & -1.03 & -1.05    \\     
\enddata
\tablenotetext{*}{Empty entries indicate that no solution exists for this $V_\di{A}$ value.}
\tablenotetext{\dagger}{Solutions with Alfv\'en singularities (see text).}
\end{deluxetable}

Eastward and westward  equatorial IG/MIG solutions are the system's most rapidly oscillating waves (with $|\omega|> c_g k$). The azimuthal background magnetic field slightly increases the phase speed of the MIG modes (see \cref{tab:eigenvalues}). However, their energy redistribution patterns remain qualitatively similar to their hydrodynamic IG counterparts (see \cref{fig:eigenfunctions}, rows one and three).

Kelvin/magneto-Kelvin solutions are characterised by zonally-dominated winds. The are two hydrodynamic Kelvin solutions: an eastward equatorial Kelvin solution, with $\omega =  c_g k$, $\hat{v}=0$, $\{ \hat{u},\hat{h} \} \propto \exp(- y^2/2L_\di{eq}$), and  a westward boundary Kelvin solution, with $\omega =  -c_g k$, $\hat{v}=0$, $\{ \hat{u},\hat{h} \} \propto \exp(y^2/2L_\di{eq}$).\footnote{The westward boundary Kelvin solution is removed  when the condition is $\{ \hat{u},\hat{v},\hat{h} \} \rightarrow 0$ as $|y| \rightarrow 0$ is imposed \citep{196625}.} These hydrodynamic solutions are special cases of magneto-Kelvin eigensolutions, which have $c_g k \leq |\omega| \leq (c_g^2 + V_\di{A}^2 \ee)^{1/2} k $. While hydrodynamic Kelvin solutions have $\hat{v}=0$ everywhere, we find that magneto-Kelvin solutions acquire a non-zero $\hat{v}$ in order to maintain latitudinally-independent oscillation frequencies. This can be understood by combining \cref{eqn:mom:lin1,eqn:cont:lin,eqn:ind:lin} to yield
\beq
(\omega^2-c_m^2 k^2) \hat{u} =  i f \omega \hat{v} -  i k c_g^2 \oderiv{\hat{v}}{y}. \label{eqn:uhat}
\eeq
For hydrodynamic Kelvin solutions, the lefthand and righthand sides of \cref{eqn:uhat} are identically zero throughout the domain; whereas magneto-Kelvin solutions have $c_g k \leq |\omega| \leq (c_g^2 + V_\di{A}^2 \ee)^{1/2} k $, $\{\hat{u}, \hat{h}\}$ similar to their hydrodynamic counterparts, and a non-zero $\hat{v}$ that ensures \cref{eqn:uhat} remains balanced. Like in the hydrodynamic limit, we find two magneto-Kelvin solutions: an eastward equatorial magneto-Kelvin solution and  a westward boundary magneto-Kelvin solution. Magnetism causes both varieties to have a small non-zero meridional velocity component ($|\hat{v}/\hat{u}|\ll1$) and an increased $|\omega|$, but both are characteristically similar to their hydrodynamical counterparts. For the equatorial magneto-Kelvin solution, this is illustrated in \cref{fig:eigenfunctions}, which shows its energy redistribution pattern remains qualitatively similar as $V_\di{A}$ is increased.

In the hydrodynamic version of the system, equatorial Rossby solutions  propagate westward and oscillate slowly ($|\omega|<c_g k$), with their azimuthal phase speeds, $|\omega|/ k$,  successively decreasing for larger $n$ solutions. In the hydrodynamic limit, the structures of equatorial Rossby solutions are characterised by mid-to-high latitude geostrophic vortices (see \cref{fig:eigenfunctions}, row two, lefthand  column). For weakly-magnetic equatorial magneto-Rossby solutions, we find that the presence of the azimuthal background magnetic field has little effect on the form of the waves' eigenfunctions, which are magnetogeostrophic in nature. Weakly-magnetic solutions adjust to the contribution of magnetic tension with small increases to their azimuthal phase speeds. However, when their oscillation frequencies are exceeded by the maximal background azimuthal \Alf frequency (i.e., when $ V_\di{A} \geq \ee^{-1/2}  |\omega|/ k$), equatorial magneto-Rossby solutions enter the $\omega$-range of \Alf singularities and are removed from the system. Higher $n$ equatorial magneto-Rossby solutions are removed for the weakest $ V_\di{A}$ values, before  successively lower $n$ solutions are removed for larger $ V_\di{A}$ values (as Alfv\'enic properties become dynamically important at larger and larger scales). We attribute the removal of the planetary scale equatorial magneto-Rossby solutions to the breaking of potential vorticity conservation in regions of large Lorentz force

The shallow-water hydrodynamic definition of potential vorticity is, $q = h^{-1}(\textderiv{v}{x}-\textderiv{u}{y} + f)$ \cite[e.g.,][]{2006aofd.book.....V}. In the non-diffusive, unforced, drag-free version of the SWMHD model, the potential vorticity evolution satisfies
\beq
\lagderiv{q} = \oo{h}[\nabla\times(\vect{J} \times \vect{B})]\cdot \uvect{z}, \label{eqn:SWMHD:PV}
\eeq
where $\vect{J}=  (\textderiv{B_y}{x}-\textderiv{B_x}{y}) \uvect{z} $. \cref{eqn:SWMHD:PV} shows that the curl of the Lorentz force generated by the horizontal magnetic field component generally prevents potential vorticity conservation in the magnetic limit.\footnote{Further, \cite{2002PhPl....9.1130D} showed that potential vorticity has no materially invariant counterpart in SWMHD.} Since the material conservation of potential vorticity is essential to the propagation mechanism of Rossby waves \cite[e.g., see][]{2006aofd.book.....V}, in regions of large Lorentz force their generation is inhibited.

In magnetic systems, two additional sets of solutions emerge. These solutions have  $ |\omega| \leq V_\di{A} \ee^{1/2} k$, so contain singularities, yet present some distinguishable properties of \Alf waves. Specifically, they arise in both eastward and westward travelling varieties, and $|\omega|$ increases with $V_\di{A}$ and $n$. To assess their nature as $y\rightarrow y_s$ (for $\omega^2 = B_0(y_s)^2 k^2$), we use the Frobenius  solutions discussed in  \appref{app:sing:solns}. In singular regions, $\hat{v}=O(\ln|(y-y_s)/L_\di{eq}|)$, 
so by \cref{eqn:uhat} $\hat{u}=O([(y-y_s)/L_\di{eq}]^{-1})$, meaning that $|\hat{u}/\hat{v}|\rightarrow \infty$ as  $y\rightarrow y_s$. This highlights that \Alf singularities cause a wave barrier to emerge at $y=y_s$, over-which wave-driven meridional energy/momentum transport mechanisms cannot cross. Since they are not finite everywhere, equatorial wave structures with \Alf singularities cannot determine global energy redistribution in the same way that planetary-scale equatorial waves do in hydrodynamic hot Jupiter models. Hence, in the limit where magnetism becomes significant, dissipative and non-linear effects become essential for understanding equatorial dynamics.  A  non-singular analogue of these solutions could be present in systems that include these extra physical processes but, since we are focused on the breakdown of geostrophic balance, we do not investigate solutions of this kind further.\footnote{In the very strong field limit, \cite{2017GApFD.111..115L} identified ``outer band'' solutions akin to these Alfvenic solutions that were trapped in polar regions in linear non-diffusive beta-plane systems, but concluded that they do not have a finite global (linear, non-diffusive) counterpart in  \cite{2018GApFD.112...62L}. Spherical (linear, non-diffusive) SWMHD waves studies in other geometries have found additional  {slow magneto-Rossby} \citep{2017GApFD.111..282M} and {magnetostrophic} \citep{2009ApJ...703.1819H} type waves at the poles of shallow-water systems, which may be useful in explaining the dynamics of the polar MHD flows. \cite{2017GApFD.111..282M} also found polar trapping of the ``fast'' magneto-Rossby solutions, which can plausibly be related to the removal of equatorial magneto-Rossby solutions (i.e., magneto-Rossby waves could become confined to regions of the atmosphere less influenced by magnetism).  }

 Thus far, we have discussed magnetic free wave solutions about a flat rest state. However, \cite{2014ApJ...793..141T} and \cite{2020A&A...633A...2D} find the redistributing properties of waves can be altered by the presence of a background zonal flow, though the fundamental characteristics of these waves remain unchanged. Compared to the system we have so far explored, taking $u_0=U_0$ constant \citep[as in][]{2014ApJ...793..141T}, simply manifests itself in the trivial phase translation $\omega \mapsto  \omega^* - U_0 k$, where $\omega$ and $\omega^*$ are oscillation frequencies for a background at rest and a background with a zonal flow respectively. For this translation, \Alf singularities emerge where $B_0(y_s)^2 k^2 = (\omega^* - U_0 k)^2=\omega^2$, which is the same condition as the rest case. We have also considered solutions about the latitudinally dependent background state, $u_0=u_0(y)$, finding that Alfv\'enic singularities, with similar Frobenius solution dependencies, emerge at points where $B_0(y_s)^2 k^2 = (\omega^* - u_0(y_s)k)^2$.

\subsection{Comparisons with non-linear simulations}
Our findings concerning Alfv\'en-Rossby wave coupling in an equatorial beta-plane model, with an equatorially-antisymmetric azimuthal background magnetic field, are consistent with our developed theory of hotspot reversals from the simulations of \sectref{sect:numerical:results}. In the hydrodynamic limit, planetary scale geostrophic circulations associated with equatorial Rossby waves are free to recirculate energy between the equatorial and mid-to-high latitudes in a manner described by \cite{2011ApJ...738...71S}. In the weakly-magnetic limit, planetary scale circulations remain largely unchanged, with equatorial magneto-Rossby waves only altering slightly to account for the magnetic contribution to their magneto-geostrophic circulations. However, at a critical threshold magnetic tension becomes large enough to inhibit the magneto-geostrophic circulations associated with equatorial magneto-Rossby waves.  This is the free wave manifestation of the obstruction of geostrophic circulations, which we identified as the trigger for hotspot reversals in \sectref{sect:numerical:results}. Here the analogy between global circulations and the standing wave description of linear steady-state solutions described by \cite{2011ApJ...738...71S} breaks down and the force balance description used in \sectref{sect:numerical:results} is preferable. In \sectref{sect:numerical:results}, we saw that the meridional Lorentz force responsible for obstructing geostrophic circulations always has a corresponding westward component that, ultimately, results in hotspot reversals. Together, the developed theory of \sectsref{sect:numerical:results}{sect:linear:model} can be used to place a zero-amplitude limit on the reversal threshold, $V_{\di{A},\di{crit}}$. In linear theory, magnetic tension inhibits the propagation of equatorial Rossby waves, with the oscillation frequency
\beq
\omega_{\mathrm{R},n} = \frac{-\beta k}{k^2 + (2n+1)\beta/c_g}, \label{eqn:omega:Rossby}
\eeq 
when $\omega_{\di{A},\di{max}} \geq |\omega_{\mathrm{R},n}| $, where $
\omega_{\di{A},\di{max}} = B_{0,\di{max}} k = V_\di{A} \ee^{1/2} k
$ is the maximal \Alf frequency. Our findings suggest that, when the slowest (largest $n$) equatorial Rossby wave that is important for supporting the planetary scale mid-to-high latitude geostrophic balance becomes inhibited by magnetic tension, geostrophic circulations are obstructed and hotspots are driven westward by the resulting zonal Lorentz force. 

 In \cref{fig:VAcrit:numeric}, we have overplotted the theoretical thresholds associated with the obstruction of the $n=1,3,5$ equatorial Rossby solutions, for comparison with the  zero-amplitude ($\Delta h_\di{eq}/H\rightarrow 0$) limits of the simulated reversal thresholds, $V_{\di{A},\di{crit}}$. We generally find acceptable agreement between the simulations and these theoretical criteria, noting that in the most physically relatable case, where $\trad$ is short, reversals occur at the point where the $n=1$ equatorial Rossby wave is overcome by magnetic tension. When $\trad$ and $\tdrag$ act over longer timescales, \cref{fig:VAcrit:numeric} suggests that the obstruction of geostrophic circulations is associated with the loss of larger $n$ equatorial Rossby solutions. This is somewhat consistent with the standing wave description of linear steady-state hydrodynamic solutions, as geostrophic circulations in solutions with longer $\trad$ and $\tdrag$ timescales are located at higher latitudes (e.g., see \cref{fig:Geop}), so require contributions to their energy recirculation patterns from larger $n$ equatorial Rossby waves \citep[e.g.,][]{196625,2011ApJ...738...71S,2014ApJ...793..141T}. While a wave analysis with non-linear effects and diffusion may be able to more precisely define these weakly-forced limits, we note that this
 description provides a vast improvement on scaling predictions of typical toroidal field strengths on hot Jupiters, which have order of magnitude (or larger) uncertainties (discussion in \sectref{sect:hotspot:reversal:criterion}).

\begin{figure*}
\centering
\includegraphics[width=0.6\linewidth]{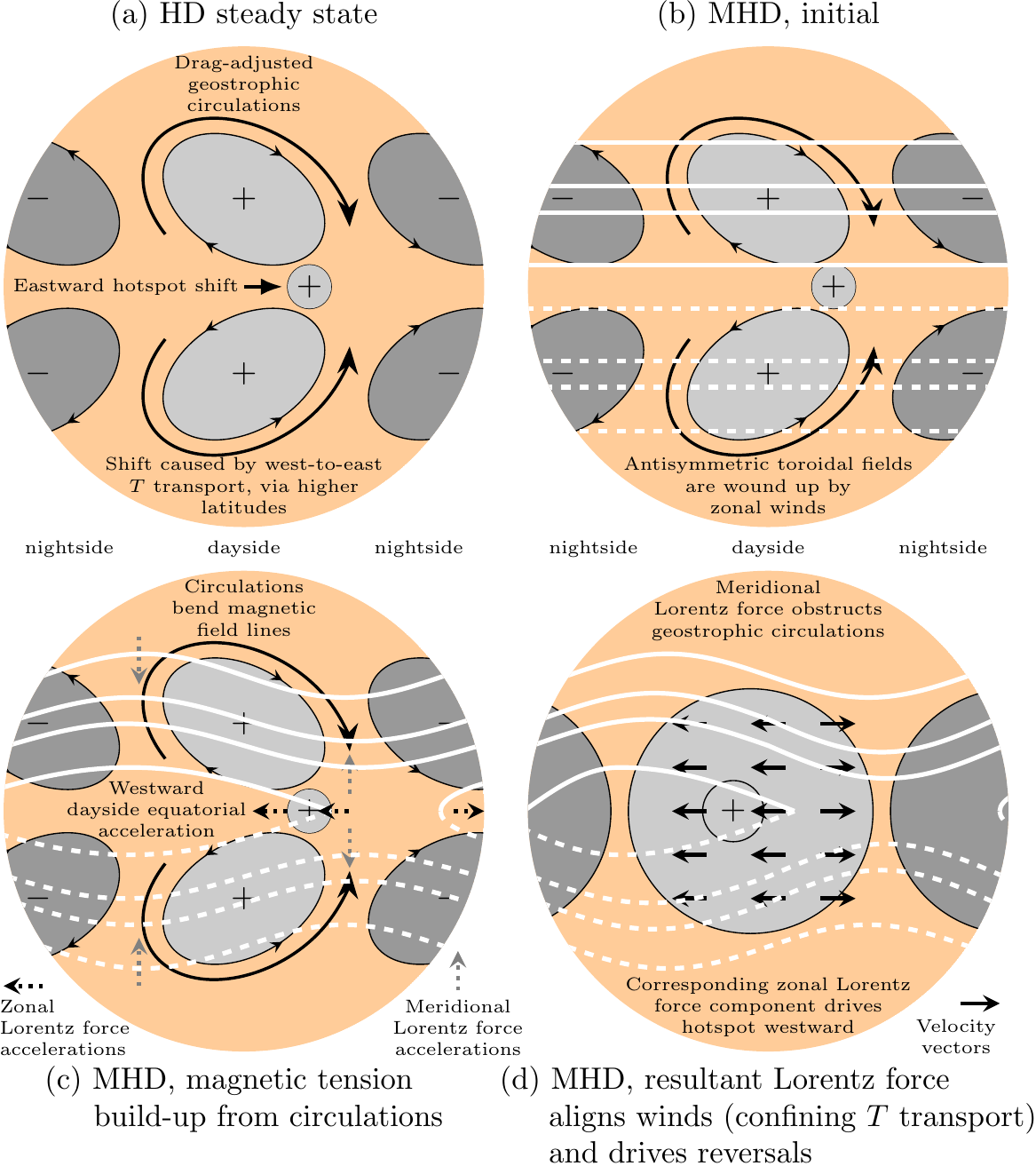} %
\caption{A schematic of the magnetic reversal mechanism, with grey temperature contours and white magnetic field lines (solid for $B_x>0$; dashed for $B_x<0$). $(a)$ In hydrodynamic steady state solutions, drag-adjusted geostrophic circulations dominate at mid-to-high latitudes; whereas zonal pressure-driven jets dominate at the equator. Hotspots are shifted eastward as these circulations transport thermal energy from the western equatorial dayside to the eastern equatorial dayside, via higher latitudes. $(b)$ In ultra-hot Jupiters, partially-ionised winds flow through the planet's deep-seated magnetic field, inducing a dominant equatorially-antisymmetric atmospheric toroidal magnetic field. When field lines are parallel to the equator, magnetic tension is zero, so flows behave hydrodynamically. $(c)$ As the field and flow couple, the geostrophic circulations bend the magnetic field lines poleward on the western dayside and equatorward on the eastern dayside, generating a Lorentz force, $(\mathbf{B}\cdot \nabla)\mathbf{B}$. The meridional Lorentz force component acts to resist the geostrophic circulations; whereas, since $|B_x|$ is smallest in equatorial regions, the zonal Lorentz force component, $(\mathbf{B}\cdot \nabla)B_x$, is westward in hotspot regions, where field lines bend equatorward (and vice versa where field lines bend poleward). $(d)$ Beyond a magnetic threshold, the system's nature changes. The meridional Lorentz force obstructs the circulating geostrophic winds, causing zonal wind alignment. This confines thermal structures and blocks the hydrodynamic transport mechanism. The zonal Lorentz force accelerates winds westward in the hottest dayside regions, causing a net westward dayside temperature flux. This drives the hottest thermal structures westward, until zonal pressure gradients can balance the zonal Lorentz force.}  \label{fig:reversal:schematic}
\end{figure*}

\section{The magnetic reversal mechanism} \label{sect:hotspot:reversal:criterion}

In \sectref{sect:numerical:results}, we identified the mechanism that drives magnetic hotspot reversals in SWMHD simulations of hot Jupiters. We provide a schematic and summarised explanation of the mechanism in \cref{fig:reversal:schematic} and its caption.

This mechanism is also relevant for other, less idealised, magnetic field geometries. The reversal mechanism requires two features in the azimuthal field geometry:
(1) large $|B_x|$ at mid-to-high latitudes to block/obstruct the circulation of the energy transporting geostrophic flows and (2)
smaller or zero $|B_x|$ at equatorial latitudes, so that when magnetic field lines are bent into the equatorial region (by the mid-to-high latitude circulations), they pass into regions of smaller $|B_x|$,  generating a westward Lorentz force acceleration. This suggests that, as long as the profile is characterised by these two features, the developed theory does not depend on exact antisymmetry in the dominant magnetic field geometry. This observation is useful when comparing to the 3D MHD simulations of \cite{2014ApJ...794..132R} and \cite{2017NatAs...1E.131R}, which are characterised by antisymmetrically-dominant, but not exactly antisymmetric, toroidal magnetic field geometries.

\subsection{Hotspot reversal criterion}
In \sectsref{sect:numerical:results}{sect:linear:model}, we identified two physically-motivated reversal criteria on the \Alf speed. The azimuthal \Alf speed is defined as $V_\di{A}= B_\phi / \sqrt{\mu_0 \rho}$, where $\mu_0$ and $\rho$ are the permeability of free space and the density. Taking $c_g = \sqrt{\mathcal{R}T}$ (see \sectref{sect:numeric:model}) and applying the ideal gas law therefore yields $  B_{\phi}  \sim (V_{\di{A}}/{c_g}) \sqrt{\mu_0 P} $,  where $T$ and $P$ are the temperature and pressure at which the reversal occurs. From this, we have the following critical reversal criterion on the toroidal field magnitude:
\beq 
 \begin{split}
 & B_{\phi,\mathrm{crit}} \approx  \sqrt{\mu_0 P}  \max\left[  \frac{\beta / c_g}{k^2 + 3\beta/c_g},   \right. 
\\ & \;\; \left. \frac{  2 \pi R}{L_\di{eq}}  \left(\frac{\Delta h_\di{eq}}{H}\right)\left( \frac{\trad}{\twave}\right)^{-1} \left(\frac{2 \Omega \twave^2}{\trad} + 1 \right)^{-1}\right], \label{eqn:Bphi:criterion}
\end{split}
\eeq
where $n=1$ (largest scale Rossby wave) and $\kappa \approx 1$ \citep[$B_{\phi}$ approaches maximal amplitudes close to the equator, as in][]{2014ApJ...794..132R}  have been taken. This criterion quantifies the toroidal field magnitude sufficient to obstruct geostrophic circulations, with the first term in the maximum relating to when the toroidal field inhibits the propagation of the largest scale equatorial Rossby wave (in the small $\Delta h_\di{eq}/H$ limit).

Further, if the electric currents that generate the planet's assumed deep-seated dipolar field are located far below the atmosphere, \cite{2012ApJ...745..138M} argued that the toroidal and dipolar field magnitudes should be related by the scaling law: $B_\phi \sim R_m  B_\di{dip}$, where  $R_m = {\mathcal{U}_\phi H}/{\eta}$ is the magnetic Reynolds number and  $\mathcal{U}_\phi$ is the magnitude of zonal wind speeds. We use the toroidal field criterion, and apply $B_\phi \sim R_m  B_\di{dip}$, to quantitively compare the predictions of SWMHD theory to the 3D MHD simulations of \cite{2014ApJ...794..132R} and \cite{2017NatAs...1E.131R}.

\subsection{Comparisons between SWMHD and 3D MHD}

\subsubsection{Linking hotspot and wind reversals}

Thus far, we have considered hotspot reversals, rather than the reversal of zonal-mean zonal winds, $\bar{u}$. Though time-correlated in 3D MHD models \citep{2017NatAs...1E.131R}, hotspot and wind reversals are not necessarily synonymous. While thermal/wind structures and geopotential/wind structures compare well between hydrodynamic shallow-water and 3D models \citep[e.g.,][]{2013ApJ...776..134P,2016ApJ...821...16K}, \cite{2020A&A...633A...2D} found a consistent treatment of the vertical component of the eddy-momentum flux (i.e., the vertical Reynolds stress) is critical to the development of equatorial superrotation ($\bar{u}>0$).\footnote{Interestingly, while SWMHD does not include a consistent treatment of vertical eddy-momentum flux, we still find that $\bar{u}$ can reverse during the transition phase (only) of hotspot reversals in supercritical SWMHD simulations.}

In hydrodynamic models of hot Jupiters, equatorial superrotation emerges from the momentum transport mechanism of \cite{2011ApJ...738...71S}. \cite{2011ApJ...738...71S} noted that the necessity for such a mechanism is a consequence of an angular momentum conservation theorem arising from \cite{1969JAtS...26..841H}, which implies that equatorial superrotation can only be maintained if driven by an up-gradient angular momentum pumping mechanism. \cite{2011ApJ...738...71S} showed that this up-gradient mechanism is provided by the same geostrophic circulations that result in eastward hotspots. Therefore, since we have shown that magnetically-driven hotspot reversals are caused by the obstruction these recirculation patterns, Hide's theorem provides an anti-theorem, which implies that the magnetically-driven hotspot reversals are accompanied by a disruption of superrotation.

The realisation of this anti-theorem can be identified in 3D MHD simulations. These found that mid-to-high latitude vortical structures zonally-align and, consequently, the transport of eastward eddy-momentum (horizontal Reynolds stress) from mid-latitudes into equatorial regions is reduced at atmospheric depths were reversals occur \citep[compare Figures 2, 9, and 11 in][]{2014ApJ...794..132R}.  \cite{2014ApJ...794..132R} found that, when the up-gradient horizontal Reynolds stress component diminishes,  westward equatorial zonal-mean zonal accelerations are driven by the remaining down-gradient momentum transport components (i.e., the vertical Reynolds stress and the Maxwell stresses). Thus, the above the application of Hide's theorem provides a meaningful connection between wind reversals and the magnetically-driven hotspot reversals mechanism we have presented.

\subsubsection{Wave dynamics and turbulence}
While we have not modelled turbulence in this work, actual planetary flows are expected to be highly turbulent. In hydrodynamic planetary systems, wave arguments have historically proved useful for developing understanding of geostrophic turbulence and how its conservational properties relate to eddies. Specifically, potential vorticity conservation is fundamental for both Rossby wave propegation and geostrophic turbulence, so Rossby wave properties can be used to understand the structures of planetary scale turbulence  \citep[e.g.,][]{rhines_1975,2006aofd.book.....V}. \cite{2014ApJ...794..132R} found that the relationship between zonal jets and magnetic fields in 3D MHD simulations shared intermittent features with MHD turbulence on a beta-plane that were identified by \cite{2007ApJ...667L.113T}. Hydrodynamic geostrophic turbulence and MHD beta-plane turbulence have very different characteristics. Amongst them, the wave-wave/wave-zonal flow interactions associated with the inverse cascade of geostrophic turbulence are replaced with interactions that result in a forward MHD cascade, with MHD interactions occurring over scales on (and below) the planetary scale when the azimuthal \Alf wave frequencies exceed the planetary scale Rossby wave frequency \citep{2007sota.conf..213D}. This turbulence condition is remarkably similar to the hotspot reversal criterion we identified in the weakly forced regime, which was motivated by wave dynamics and the findings of non-turbulent SWMHD simulations. We attribute this kinship to the breaking of potential vorticity conservation in MHD models in regions of large horizontal Lorentz force, which inhibits  geostrophic characteristics such as Rossby wave propagation  (as discussed in \sectref{sect:linear:model}). 
We also highlight that forcing and drags generate potential vorticity sources/sinks, so potential vorticity conservation is modified when drag and forcing treatments are strong, which is why  reversal thresholds deviate from this simple criterion in the strongly forced limit. 

\subsubsection{Magnetic field evolution and structure}
After the initial hotspot transition, long term temporal differences between SWMHD and  3D MHD models arise because SWMHD can only model the planetary dipolar field {\em or} the atmospheric toroidal field self-consistently (see \sectref{subsect:initial:Bfield}), meaning that it cannot take into account toroidal field induction from reversed conducting zonal winds passing through the planetary dipolar field. If a strong toroidal field can be maintained indefinitely, the shallow-water theory predicts completely reversed winds, even in 3D models. However, at the onset of the wind reversals, the induction caused by the reversed winds flowing  through the deep-seated magnetic field will result in a reduction of the atmospheric toroidal field's magnitude. Hence, while the quasi-steady magnetically-driven wind reversals of SWMHD are useful for modelling the reversal process, in reality one would expect to see oscillatory wind variations as toroidal fields successively strengthen and weaken in a wind-up-wind-down cycle of the toroidal magnetic field. Wind variations of this kind can be both observationally inferred from the oscillating peak brightness offsets of HAT-P-7b \citep{2016NatAs...1E...4A} and directly measured in 3D MHD simulations of the HAT-P-7b parameter space \citep{2017NatAs...1E.131R}. This in itself has the interesting consequence that the reversal mechanism may provide a saturation process for the atmospheric toroidal magnetic field. 

Due to the density dependence of the \Alf speed,  $B_{\phi,\mathrm{crit}}$ has a $\sim P^{1/2}$ pressure dependence (see \cref{eqn:Bphi:criterion}). This explains why \cite{2014ApJ...794..132R} and \cite{2017NatAs...1E.131R}
found that wind reversals first onset in the upper atmosphere, but move deeper for stronger field strengths. Furthermore, if 
the reversal mechanism is a toroidal field saturation process (as discussed above), $B_\phi$ should not  greatly exceed $B_{\phi,\di{crit}}$. Hence, 
$B_\phi$ should decrease above the deepest region where reversals occur (since $B_{\phi,\di{crit}}$ decreases upwards), which is a feature of the toroidal field profiles found in \cite{2014ApJ...794..132R}, though other processes may also cause an upwards reduction in $B_\phi$. Comparing the geometry of the toroidal fields in the quasi-steady reversed SWMHD solutions with those in oscillating 3D MHD solutions is difficult. However, when the toroidal field is approaching criticality in strength, we do find similarities between our toroidal field geometries and those of \cite{2014ApJ...794..132R}. In both models the equatorially-antisymmetric toroidal fields couple to mid-to-high latitude circulations in a manner that bends them towards the equator from west to east, which we showed is a geometry that results in westward Lorentz force accelerations (see \sectref{sect:numerical:results}).

\begin{deluxetable*}{C|c|C|C|C|C|C|C}
\tablecaption{Estimates of reversal criteria compared to field strengths and reversal criteria from 3D MHD simulations (see \sectref{subsubsect:3D:comp} for definitions and an accompanying discussion).  \label{tab:3D:comp}} 
\tablehead{
\colhead{Model} & 
\colhead{Parameters} & 
\colhead{$P/\mbar$}  & 
\colhead{$T_\di{eq}/\kelvin$}  & 
\colhead{$\Delta T /T_\di{eq}$} &
\colhead{$B_{\phi,\di{crit}}/\gauss$}   & 
\colhead{$B_{\di{dip},\di{crit},\di{base}}/\gauss$}  &
\colhead{3D comparison}      
}   
\startdata
\mbox{M7b1}^a & HD209458b    & 20^*  & 1850  & 0.1\mbox{-}0.2  & 175\mbox{-}350& - & |B_{\phi}|=220 \gauss  \\
\mbox{M7b2}^a & HD209458b    & 200^*  & 1950 & 0.05\mbox{-}0.1 & 430\mbox{-}545 & - & |B_{\phi}|=510 \gauss   \\    
\mbox{M7b2}^a & HD209458b    & 10^\dagger  & 1750 & 0.15\mbox{-}0.2 & 185\mbox{-}247 & - & |B_{\phi}|=190 \gauss   \\    
\mbox{HAT-P-7b}^b & HAT-P-7b &  1^* &  2200  & 0.22& 92 & 7 &  3\gauss<B_{\di{dip},\di{crit},\di{base}}<10\gauss\\      
\enddata
\tablenotetext{*}{Critical wind reversal depth, $P \approx P_\di{crit}$. $^\dagger$ Above critical wind reversal depth, $P<P_\di{crit}$.}
\tablerefs{$^a$\cite{2014ApJ...794..132R}; $^b$\cite{2017NatAs...1E.131R}}
\end{deluxetable*}

 \subsubsection{Quantitive comparisons with 3D MHD} \label{subsubsect:3D:comp}

In \cref{tab:3D:comp}, we compare predictions of the reversal criterion to magnetic field strengths of in three 3D MHD simulations:  M7b1 and M7b2 of \cite{2014ApJ...794..132R}, and the HAT-P-7b model of \cite{2017NatAs...1E.131R}, all of which display wind reversals at some critical pressure depth, $P_\di{crit}$. In these estimates, we take $T_\di{eq} = \bar{T}$, $\Delta T = T_\di{day}-\bar{T}$, $\trad=\twave$, and set $\Delta h_\di{eq}/H = \Delta T/T_\di{eq}$. For comparisons to the simulations of \cite{2014ApJ...794..132R}, we compare $B_{\phi,\di{crit}}$ to $|B_{\phi}|$,  the horizontally-averaged toroidal field component at the end of the run; whereas, for the HAT-P-7b  simulation of \cite{2017NatAs...1E.131R}, we estimate the critical dipolar field strength at the atmospheric base, $B_{\di{dip},\di{crit},\di{base}}$. This is calculated using the $B_\phi \sim R_m  B_\di{dip}$ scaling law of \cite{2012ApJ...745..138M}. We take $\eta= \num{2e6} \,\mbox{m}^2\, \mbox{s}^{-1} $ and  $\mathcal{U}_\phi \sim 10^2 \,\mbox{m}\,\mbox{s}^{-1}$ from 3D simulations, to yield $B_{\di{dip},\di{crit}}\approx 4.3\gauss$ at $P=1\mbar$, then noting that the atmospheric base is located at $r=0.15 R$ in the simulations yields $B_{\di{dip},\di{crit},\di{base}}=7$. We note that the reversal criterion compares reasonably to the magnitude of the horizontally-averaged toroidal component field in the simulations of \cite{2014ApJ...794..132R}, with uncertainties in $T_\di{day}$ bracketing the true $|B_{\phi}|$ value. This occurs both at $P=P_\di{crit}$ and above $P_\di{crit}$, supporting the idea of reversals providing a toroidal field saturation process. The prediction of  $B_{\di{dip},\di{crit},\di{base}}=7$ lies within the range $3\gauss<B_{\di{dip},\di{crit},\di{base}}<10\gauss $ identified by \cite{2017NatAs...1E.131R}. We note that, while $B_{\phi,\di{crit}}$ has dependencies on $\Delta T/T_\di{eq}$ and $\trad$,  $\eta$ can vary significantly between the day and night sides of ultra-hot Jupiters (by orders of magnitude). Therefore, current understanding of the connection between toroidal and poloidal fields on hot Jupiter is constrained by large uncertainties (in $ B_\phi \sim R_m  B_\di{dip}$), which far outweigh uncertainties in the toroidal field criterion that we have developed.

\section{Discussion} \label{sect:discussion}

In this work we have explained the atmospheric mechanics of magnetically-driven hotspot reversals in hot Jupiters using  numerical (\sectref{sect:numerical:results}) and semi-analytic (\sectref{sect:linear:model}) analyses of a SWMHD model (\sectref{sect:numeric:model}), where we have applied parameters based on the ultra-hot Jupiter HAT-P-7b. In \sectref{sect:hotspot:reversal:criterion} we used the theory developed throughout this study to identify a criticality criterion and discussed our findings in the context of 3D MHD simulations. This criticality criterion can be used to place physically-motivated constraints on the magnetic fields of ultra-hot Jupiters with observed westward hotspots. It also represents the point where hydrodynamic models with Lorentz force mimicking Rayleigh drag treatments should be replaced with self-consistent MHD modelling. In \sectref{sect:hotspot:reversal:criterion}, we also identified the link between wind reversals and the hotspot reversal mechanism, highlighted relevant shared features between modifications to wave dynamics and atmospheric turbulence, discussed the role of reversals on the magnetic field's evolution, and made quantitive comparisons between the reversal criterion 3D MHD simulations. 

Using the numerical SWMHD simulations, we demonstrated that hotspot reversals occur when equatorially-antisymmetric azimuthal components of the magnetic field are strong enough to obstruct the geostrophic circulations that transport energy to the eastern dayside in hydrodynamic models. The magnetic field geometry that results from this obstruction always drives westward Lorentz force accelerations in hotspot regions, causing hotspots to transition from east-to-west. Using this finding we identified a reversal criterion for the toroidal field in the strong forcing regime using a simple argument based on the timescales of the two competing processes.

The recent observational drive in exoplanet meteorology provides a timely backdrop around which theories regarding the mechanism of wind/hotspot reversals can be tested and developed. Observational constraints on atmospheric properties continue to improve whilst a combination of archival data and dedicated observational missions from Kepler, Spitzer, Hubble, TESS, CHEOPS (and in the future JWST) are accelerating our understanding of the atmospheric theory of exoplanets.  Since the prediction of magnetically-driven wind variations in hot Jupiters \citep{2014ApJ...794..132R}, westward hotspots/brightspots have been inferred on the hot Jupiters HAT-P-7b \citep{2016NatAs...1E...4A}, CoRoT-2b \citep{2018NatAs...2..220D}, Kepler-76b \citep{Jackson_2019}, WASP-33b \citep{2020arXiv200410767V}, and  WASP-12b \citep{2019MNRAS.489.1995B}. In \cite{2019ApJ...872L..27H} we inferred that the observed westward offsets of CoRoT-2b are unlikely to be driven by magnetism and \cite{2018NatAs...2..220D} proposed that such observations of CoRoT-2b could be explained by nonsynchronous rotation. However, as HAT-P-7b, Kepler-76b, WASP-33b, and WASP-12b are all ultra-hot Jupiters, the westward hotspots/brightspots observations on these planets are likely to be driven by magnetism. In future work, we will estimate the magnetic field strengths sufficient to explain their westward hotspots/brightspots observations.

The toroidal field hotspot/wind reversal criterion we have developed is observationally motivated and appears to reproduce results of 3D MHD simulations. While this criterion does come with uncertainties due to the simplifications we have made, there are currently much larger uncertainties in the $ B_\phi \sim R_m  B_\di{dip}$ magnitude scaling of \cite{2012ApJ...745..138M}, which is used to connect the poloidal-toroidal magnitudes. This is because $\eta$ is highly temperature dependent \citep[e.g.,][]{2014ApJ...794..132R}, meaning that $R_m$ can vary by orders of magnitude between sides of the same hot Jupiter. Three-dimensional models can further inform about connections between the poloidal-toroidal fields, which cannot be studied with SWMHD. In particular, we highlight the need to develop theoretical understanding of the effects that accompany strong day-night $\eta$ dependencies in hot Jupiter atmospheres. Such comparisons offer important testcases for the extension of dynamo theory into the hot Jupiter regime. The predictions and constraints in this paper are clearly not the end of the story and, ultimately, bespoke 3D MHD simulations offer the best prospect for providing accurate constraints on the magnetic field strengths of ultra-hot Jupiters. That said, understanding the reversal mechanism is an important theoretical step and the reversal criteria we have presented enables modellers/observers to gain intuition into the most important atmospheric characteristics concerning reversals, particularly if one is comparing between multiple hot Jupiters in an ensemble approach.

Further details of the concepts, models, and applications discussed in this work are included in A.~W.~Hindle's forthcoming PhD thesis.

\acknowledgments
We acknowledge support from STFC for A.~W.~Hindle's studentship (ST/N504191/1). T.~M.~Rogers and P.~J.~Bushby acknowledge the Leverhulme grant RPG-2017-035. We thank Andrew Gilbert, Andrew Cumming, Natalia G\'omez-P\'erez, and Toby Wood for useful conversations leading to the development of this manuscript.

\appendix

\section{Linearised steady state solutions}\label{app:forced:shooting}
To solve the linearised, non-diffusive, steady state SWMHD system considered in \sectref{sect:wave:contributions} (i.e., \crefrange{eq:SS:mom:x}{eq:SS:ind}), we reduce the system to a single inhomogenenous ordinary differential equation of the form
\beq
\mathcal{L}\{ \hat{v} \} \equiv F_1(y) \osecderiv{\hat{v}}{y} + F_2(y) \oderiv{\hat{v}}{y} + F_3(y) \hat{v} = \mathcal{Q}(y), \label{eqn:SS:ODE:app}
\eeq
where $F_1(y)$, $F_2(y)$, and $F_3(y)$ are latitudinally dependent coefficient functions, $\mathcal{L}$ is the system's second order differential operator, and $\mathcal{Q}(y)$ is the system's source term We have omitted the exact dependencies of $F_1(y)$, $F_2(y)$, $F_3(y)$, $\mathcal{Q}(y)$ for steady forced solutions (due to their cumbersome forms). These can be provided upon reasonable request. If $S(y)$ and $u_0(y)$ are symmetric about the equator and $B_0(y)$ is antisymmetric about the equator,  $\mathcal{L}$ and  $\mathcal{Q}$ are respectively symmetric and antisymmetric about the equator.

Solutions of \cref{eqn:SS:ODE:app}  on $-L_y<y<L_y$ are obtained by noting that, since $\mathcal{L}$ and  $\mathcal{Q}$ are respectively symmetric and antisymmetric about the equator,  inhomogeneous solutions are antisymmetric (i.e., $\hat{v}$ antisymmetric and $\hat{u}, \hat{h}, \hat{A}$ symmetric). Consequently, we solve \cref{eqn:SS:ODE:app} in the upper-half domain, $0<y<L_y$, with  $\hat{v}(L_y)=0$ (impermeability) and $\hat{v}(0) = 0$ (antisymmetry), before reflecting solutions. This reduced boundary value problem is solved by inverting the tridiagonal matrix that corresponds to \eqnref{eqn:SS:ODE:app} with finite difference discretizations. We fix the equatorial boundary condition $\hat{v}(0) = 0$ and vary $\dd \hat{v}/\dd y|_{y=0}$ in order to satisfy $\hat{v}(L_y)=0$, converging upon $\dd \hat{v}/\dd y|_{y=0}$ with the complex equivalent of the bracketed Newton-Raphson method discussed in \cite{1992nrfa.book.....P}.

\section{Singular test solutions in the linear equatorial wave solving method} \label{app:sing:solns}

For $V_\di{A} > 0$, we examine the nature of the test solutions of \cref{eqn:wave:ODE} in the upper-half domain, $0<y<L_y$, about the singular points, $y=y_s$. For $|\omega| \leq B_{0,\di{max}} k$, $y_s$ is located where $B_0(y_s) k  = |\omega|$ (\Alf singularities); whereas for $c_g k \leq |\omega| \leq (c_g^2 + B_{0,\di{max}}^2)^{1/2}k$, $y_s$ is located where $B_0(y_s) k = (\omega^2 - c_g^2 k^2)^{1/2}$ (magneto-gravity singularities).  The method of Frobenius gives
\beq
\hat{v} =  C_1 \hat{v}_1 + C_2 \hat{v}_2, \qquad \hat{v}_1 =  \sum_{n=0}^\infty a_n \hat{y}^{n+\mu_1}, \qquad  \hat{v}_2 =     D \hat{v}_1 \ln|\hat{y}| +   \sum_{n=0}^\infty b_n \hat{y}^{n+\mu_2};  \label{eqn:Frob:soln:all}
\eeq
where $\hat{y} = (y-y_s)/L_\di{eq}$, $C_1$ and $C_2$ are the constants of integration, $\hat{v}_1$ and $\hat{v}_2$ are the first and second fundamental solutions, $a_n$, $b_n$ and $D$ are constant coefficients to be set or determined, and
 $\mu_1\in \mathbb{Z}$ and $\mu_2\in \mathbb{Z}$ are the roots of the indicial equation given by  \cref{eqn:wave:ODE}.

About magneto-gravity singular points, $\mu_1 = 2$ and $\mu_2 = 0$, so
\beq
\hat{v} =  C_1 \sum_{n=0}^\infty a_n \hat{y}^{n+2} + C_2 \left( \sum_{n=0}^\infty b_n \hat{y}^{n} + D \ln|\hat{y}| \sum_{n=0}^\infty a_n \hat{y}^{n+2} \right),
\eeq
where one is free to set $a_0 = 1$, $b_0 = 1$, $b_2 = 0$ (in fact, or $b_2$ can be set to any constant), and use \cref{eqn:wave:ODE} to determine $D$,  $a_n$, and $b_n$. Solutions of this kind are not singular at $y=y_s$, so magneto-gravity singularities are in fact false singularities where the solution remains finite as $y\rightarrow y_s$.

About \Alf singular points, $\mu_1 = 0$ and $\mu_2 = 0$, so
\beq
\hat{v} =  C_1 \sum_{n=0}^\infty a_n \hat{y}^{n} + C_2 \left( \sum_{n=0}^\infty b_n \hat{y}^{n} + D \ln|\hat{y}| \sum_{n=0}^\infty a_n \hat{y}^{n} \right),
\eeq
where one is free to set $a_0 = 1$, $b_1 = 1$, $b_0 = 0$ (again, or $b_0$ can be set to any constant), and use \cref{eqn:wave:ODE} to determine $D$, $a_n$, and $b_n$. Solutions of this kind are dominated by the $\hat{v} = O(\ln|\hat{y}|)$
component as $y\rightarrow y_s$, so solutions with \Alf singularities have infinite discontinuities for $D\neq0$ (which we always find).

\bibliography{SWMHD_bib.bib}

\end{document}